\documentclass[12pt]{article}
\usepackage{graphicx,amsmath}

\newlength{\dinwidth}
\newlength{\dinmargin}
\renewcommand{\vec}[1]{\boldsymbol{#1}}
\newcommand{\dif}{\mathrm{d}}
\newcommand{\diff}[1]{\frac{\mathrm{d}#1}{#1}}
\newcommand{\xB}{x_{\scriptscriptstyle{B}}}
\newcommand{\slashed}[1]{\makebox[0pt][l]{/}#1}

\begin{document}
\titlepage
\begin{flushright}
  IPPP/03/29   \\
  DCPT/03/58   \\
  31 July 2003 \\
\end{flushright}

\vspace*{0.5cm}

\begin{center}

  {\Large \bf Unintegrated parton distributions\\[1ex] and inclusive jet production at HERA}

  \vspace*{1cm}

  \textsc{G. Watt$^a$, A.D. Martin$^a$ and M.G. Ryskin$^{a,b}$} \\

  \vspace*{0.5cm}

  $^a$ Institute for Particle Physics Phenomenology, University of Durham, DH1 3LE, UK \\
  $^b$ Petersburg Nuclear Physics Institute, Gatchina, St.~Petersburg, 188300, Russia

\end{center}

\vspace*{0.5cm}

\begin{abstract}
We describe how unintegrated parton distributions can be calculated from conventional integrated distributions. We extend and improve the `last-step' evolution approach, and explain why doubly-unintegrated parton distributions are necessary.  We generalise $k_t$-factorisation to $(z,k_t)$-factorisation.  We apply the formalism to inclusive jet production in deep-inelastic scattering, mainly at leading order, but we also study the extension to next-to-leading order.  We compare the predictions with recent HERA data.
\end{abstract}

\section{Introduction}

Conventionally, hard processes at proton colliders are described in terms of scale-dependent parton distributions $a(x,\mu^2)$, where $a=x\,g$ or $x\,q$.  These distributions correspond to the density of partons in the proton with (plus) momentum fraction\footnote{The plus and minus components of a parton with 4-momentum $k$ are $k^\pm\equiv k^0\pm k^3$.  In the infinite momentum frame, the plus momentum fraction $x\equiv k^+/p^+$ becomes the longitudinal momentum fraction of a proton with 4-momentum $p$.} $x$, integrated over the parton transverse momentum up to $k_t=\mu$. They satisfy DGLAP evolution in the factorisation scale $\mu$.  However, for exclusive processes, parton distributions unintegrated over $k_t$ are more appropriate.  The unintegrated distributions, $f_a(x,k_t^2,\mu^2)$, have the advantage that they exactly correspond to the quantity which enters the Feynman diagrams and therefore allow for the true kinematics of the process at small $x$, even at leading order (LO).  We will explain how the exact kinematics may be restored for general values of $x$.

The distributions depend on two hard scales, $k_t$ and $\mu$, and so the evolution is much more complicated.  For example, the gluon distribution $f_g(x,k_t^2,\mu^2)$ satisfies the CCFM evolution equation \cite{CCFM} based on angular ordering of gluon emissions along the chain, in the approximation where only the $1/z$ and $1/(1-z)$ singular terms of the splitting function $P_{gg}(z)$ are kept.  So far, working with this equation has only proved possible with Monte Carlo generators \cite{MC}.

However, in Ref.~\cite{Kimber:2001sc,Kimber:1999xc} it was shown that it is possible to obtain the two-scale unintegrated distributions, $f_a(x,k_t^2,\mu^2)$, using single-scale evolution equations for $h_a(x,k_t^2)$ with the dependence on the second scale $\mu$ introduced only at the \emph{last step} of the evolution.  We call this the KMR procedure.\footnote{An alternative formalism was given in Ref.~\cite{Kimber:2000bg}.}  Two alternatives for the evolution of $h_a(x,k_t^2)$ were considered: (i) pure DGLAP evolution and (ii) a unified evolution equation \cite{Kwiecinski:1997ee} which embodies both the leading $\log k_t^2$ (DGLAP) and $\log 1/x$ (BFKL) effects, as well as including a major part of the sub-leading $\log 1/x$ contributions.  As expected, the gluon and sea quark distributions, $f_a(x,k_t^2,\mu^2)$, extended into the $k_t>\mu$ region, and indeed populated this domain more and more as $x$ decreased.  An interesting result was that the unintegrated distributions obtained via the unified evolution of prescription (ii) were not very different from those based on the simpler DGLAP evolution of (i).  It was concluded that the imposition of the angular-ordering constraint in the last step of the evolution was more important than including the BFKL effects.  Here, we pay particular attention to probing the unintegrated quark distribution at larger values of $x$, so prescription (i) will certainly be a good approximation.

We refine and extend the KMR last-step procedure \cite{Kimber:2001sc} for determining the unintegrated parton distributions. First we note that in Ref.~\cite{Kimber:2001sc} angular ordering was imposed on both quark and gluon emissions; we correct this and only impose angular ordering on gluon emissions.  Second, the KMR procedure was based on $k_t$-factorisation \cite{ktfact} or the semihard approach \cite{semihard} (for a review, see Ref.~\cite{Andersson:2002cf}) in which the unintegrated parton distribution is convoluted with an off-shell partonic cross section where the incoming parton has virtuality $-k_t^2$.  This is only valid for gluons in the high-energy approximation where $z\to 0$, with $z$ the fraction of the (plus) momentum of the parent parton carried by the unintegrated parton.  Here, we generalise the notion of $k_t$-factorisation and show that it is more accurate to calculate observables using doubly-unintegrated distributions $f_a(x,z,k_t^2,\mu^2)$, where the parton now has virtuality $-k_t^2/(1-z)$.

\begin{figure}
  \begin{center}
    Inclusive jet production in DIS at LO\\[5mm]
    \includegraphics[width=\textwidth]{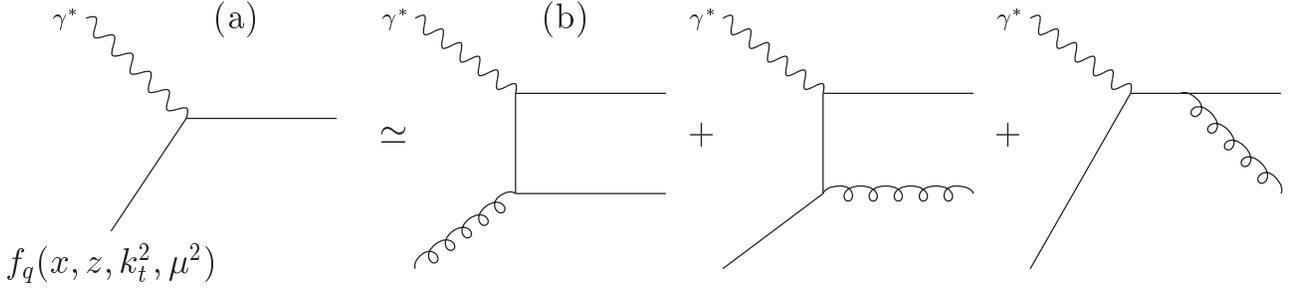}
    \caption{A schematic diagram of inclusive jet production in DIS at LO which shows the approximate equality between, on the left-hand-side (a), the formalism based on the doubly-unintegrated quark distribution, and on the right-hand-side (b), the conventional QCD approach using integrated parton densities, $a(x,\mu^2)$. \label{fig:LOjets}}
  \end{center}
\end{figure}

In Section \ref{sec:defunint} we describe how the unintegrated parton distributions, $f_a(x,k_t^2,\mu^2)$, can be determined from the conventional integrated distributions $a(x,\mu^2)$.  Then in Section \ref{sec:newfact} we define the doubly-unintegrated distributions, $f_a(x,z,k_t^2,\mu^2)$, and show how $k_t$-factorisation is generalised to $(z,k_t)$-factorisation.  The most direct way to test the unintegrated parton distributions is via inclusive jet production in deep-inelastic scattering (DIS).  Inclusive jet production, particularly in the current jet region, probes the unintegrated quark distribution in a similar way that inclusive DIS probes the integrated quark densities.  The idea is that the LO diagram computed using $(z,k_t)$-factorisation will reproduce, to a good approximation, the results of the conventional LO QCD diagrams computed using collinear factorisation.  This approximate equality is shown schematically in Fig.~\ref{fig:LOjets}.  The respective formalisms are presented in Section \ref{sec:jetprod} and their predictions for inclusive jet production are compared with each other, and also with recent HERA data, in Section \ref{sec:cfdata}.  These sections not only compare the LO predictions, but also extend the comparisons to next-to-leading order (NLO).  Section \ref{sec:conclusions} contains our conclusions.

\section{Angular-ordered parton evolution} \label{sec:evolution}

\begin{figure}
  \begin{center}
    \includegraphics[width=0.35\textwidth]{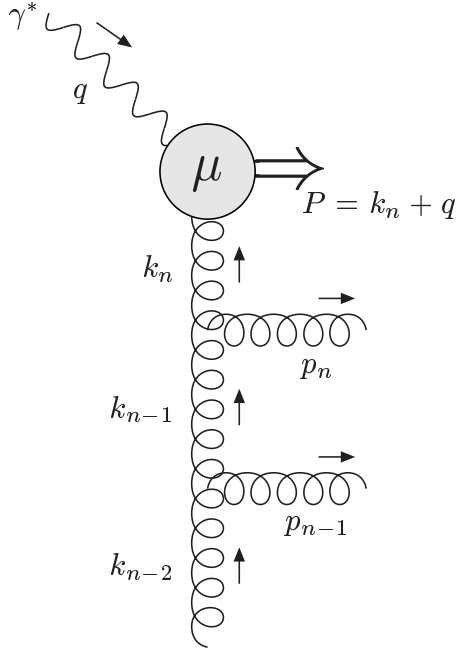}
    \caption{Upper part of the evolution chain. \label{fig:evchain}}
  \end{center}
\end{figure}

We adopt a physical (axial) gauge, which sums over only the transverse gluon polarisations, so that the ladder-type diagrams dominate the evolution.  Consider the evolution chain, illustrated in Fig.~\ref{fig:evchain}, simplified so that all the partons in the chain are gluons.  The top of the diagram indicates some hard process with a factorisation scale $\mu$.  It is convenient to use a Sudakov decomposition of the 4-momenta of the propagator gluons:
\begin{equation} \label{eq:a1}
  k_i = x_i\,p - \beta_i\,q^\prime + {k_i}_\perp,
\end{equation}
where $p$ is the 4-momentum of the proton and $q^\prime \equiv q+\xB\,p$ is a light-like 4-vector.  Here, $q$ is the 4-momentum of the photon, with virtuality $q^2=-Q^2$, and $\xB\equiv Q^2/2p\cdot q$ is the Bjorken-$x$ variable.  We neglect the proton mass since $m_p^2\ll Q^2$. In the Breit frame
\begin{equation} \label{eq:orig3}
  p = (p^+,p^-,\vec{p_t}) = (Q/\xB,0,\vec{0}), \qquad q^\prime = (0,Q,\vec{0}) \qquad \mathrm{and} \qquad {k_i}_\perp = (0,0,\vec{k_{i,t}}).
\end{equation}
The emitted gluons along the chain have 4-momenta
\begin{equation} \label{eq:a2}
  p_i = k_{i-1} - k_i = (x_{i-1} - x_i)\,p + (\beta_i - \beta_{i-1})\,q^\prime + {p_i}_\perp,
\end{equation}
while the total 4-momentum going into the hard subprocess is
\begin{equation} \label{eq:a4}
  P\equiv k_n + q = (x_n - \xB)\,p + (1-\beta_n)\,q^\prime + {k_n}_\perp.
\end{equation}
Since the outgoing partons must be on-shell ($p_i^2 = 0$), we have
\begin{equation} \label{eq:a3}
  (\beta_i-\beta_{i-1}) = \frac{\xB}{x_{i-1}(1-z_i)}\frac{p_{i,t}^2}{Q^2}\,,
\end{equation}
where $z_i \equiv x_i/x_{i-1}$, and the Sudakov (light-cone) variables of the propagator gluons obey the ordering:
\begin{equation}
  \ldots > x_{n-1} > x_n > \xB \qquad\mbox{and}\qquad\ldots <\beta_{n-1}<\beta_{n} < 1.
\end{equation}

Colour coherence effects impose the angular ordering of the gluons emitted from the evolution chain, originating from the destructive interference between the gluon emission amplitudes.  The angle between the direction of the emitted gluons, with 4-momentum $p_i$, and the proton beam direction must increase as we move towards the hard subprocess at the top of the evolution ladder. It is convenient to introduce a variable $\xi_i\equiv p_i^-/p_i^+$. Then the rapidities of the emitted gluons are
\begin{equation} \label{eq:a5}
  \eta_i=-\frac{1}{2}\log\xi_i=-\log(\tan(\theta_i/2)),
\end{equation}
and the angular ordering,
\begin{equation} \label{eq:a6}
  \ldots < \theta_{n-1} < \theta_{n} < \Theta,
\end{equation}
is equivalent to an ordering in $\xi_i$,
\begin{equation} \label{eq:a7}
  \ldots < \xi_{n-1} < \xi_{n} < \Xi,
\end{equation}
where
\begin{equation} \label{eq:a8}
  \Xi\equiv P^-/P^+=\frac{(1-\beta_n)}{x_n/\xB-1}
\end{equation}
provides the maximum allowed angle $\Theta$ via $\sqrt{\Xi}=\tan(\Theta/2)$, assuming $P^2=0$.  From \eqref{eq:a2} and \eqref{eq:a3},
\begin{equation} \label{eq:a9}
  \xi_i = \frac{p_i^-}{p_i^+} = \left(\frac{\xB\,p_{i,t}/Q}{x_{i-1}(1-z_i)}\right)^2 = \left(\frac{\xB\,\overline{p}_i}{x_{i-1}Q}\right)^2,
\end{equation}
where we have defined the rescaled {\em transverse} momenta $\overline{p}_i$ of the emitted gluons to be
\begin{equation} \label{eq:a10}
  \overline{p}_i\equiv \frac{p_{i,t}}{1-z_i}=\frac{x_{i-1}}{\xB}Q\sqrt{\xi_i}.
\end{equation}
In angular-ordered (or CCFM) evolution, the factorisation scale $\mu$ plays the r\^{o}le of the maximum rescaled transverse momentum, so $\mu=x_n Q\sqrt{\Xi}/\xB$.  Therefore, the angular ordering \eqref{eq:a7} can be written as
\begin{equation} \label{eq:angord}
  \ldots \quad z_{n-1}\overline{p}_{n-1} < \overline{p}_n\quad{\rm and}\quad z_n\overline{p}_n<\mu.
\end{equation}

\section{Unintegrated from integrated parton distributions} \label{sec:defunint}

It is informative to review how unintegrated parton distributions, $f_a(x,k_t^2,\mu^2)$, may be calculated from the conventional (integrated) parton densities, $a(x,\mu^2) = x\,g(x,\mu^2)$ or $x\,q(x,\mu^2)$, in the case of pure DGLAP evolution. Recall that the number of partons in the proton with (plus) momentum fraction between $x$ and $x+\dif x$ and transverse momentum $k_t$ between zero and the factorisation scale $\mu$ is
\begin{equation} a(x,\mu^2)\,\diff{x}\,, \end{equation}
whereas the number of partons with (plus) momentum fraction between $x$ and $x+\dif x$ and transverse momentum squared between $k_t^2$ and $k_t^2+\dif k_t^2$ is
\begin{equation} f_a(x,k_t^2,\mu^2)\,\diff{x}\,\diff{k_t^2}. \end{equation}
Thus the unintegrated distributions must satisfy the normalisation relation,
\begin{equation} \label{eq:norm}
  a(x,\mu^2) = \int_0^{\mu^2}\!\diff{k_t^2}\,f_a(x,k_t^2,\mu^2).
\end{equation}
The KMR proposal \cite{Kimber:2001sc} to determine the unintegrated distributions was to relax the DGLAP strong ordering in the last evolution step only, that is, $\ldots \ll {k_{n-1,t}} \ll k_t \sim \mu$, where we have omitted the subscript $n$ on the $k_t$ of the last propagator. This procedure is expected to account for the major part of the conventional next-to-leading logarithmic (NLL) terms, that is, terms like $\alpha_S(\alpha_S \log\mu^2)^{n-1}$, compared to the usual leading logarithmic approximation (LLA) where only terms like $(\alpha_S \log\mu^2)^n$ are included. The procedure is as follows. We start from the LO DGLAP equation evaluated at a scale $k_t$:
\begin{equation} \label{eq:DGLAP}
  \frac{\partial\, a(x,k_t^2)}{\partial \log k_t^2}  = \frac{\alpha_S(k_t^2)}{2\pi}\,\sum_{b=g,q} \left[\int_x^1\!\dif{z}\,P_{ab }(z)\,b  \left(\frac{x}{z},k_t^2 \right) - a(x,k_t^2)\int_0^1\!\dif{\zeta}\;\zeta \,P_{b  a}(\zeta ) \right ],
\end{equation}
where $P_{ab}(z)$ are the unregulated LO DGLAP splitting kernels. The two terms on the right hand side correspond to real emission and virtual contributions respectively.  The extra factor of $\zeta$ in the virtual term avoids double-counting the $s$- and $t$-channel partons.  The factor $\zeta$ is equivalent to a factor of a half when integrating over $\zeta$ and summing over $b$.

The virtual (loop) contributions may be resummed to all orders by the Sudakov form factor,
\begin{equation} \label{eq:Sudakov}
  T_a (k_t^2,\mu^2) \equiv \exp \left (-\int_{k_t^2}^{\mu^2}\!\diff{\kappa_t^2}\,\frac{\alpha_S(\kappa_t^2)}{2\pi}\,\sum_{b}\,\int_0^1\!\dif{\zeta}\;\zeta \,P_{b  a}(\zeta ) \right ),
\end{equation}
which gives the probability of evolving from a scale $k_t$ to a scale $\mu$ without parton emission.  Differentiating, we obtain
\begin{equation}
  \frac{1}{T_a (k_t^2,\mu^2)}\frac{\partial\,T_a (k_t^2,\mu^2)}{\partial \log k_t^2}=\frac{\alpha_S(k_t^2)}{2\pi} \sum_{b } \int_0^1\!\dif{\zeta }\;\zeta \,P_{b  a} (\zeta ), 
\end{equation}
so that the DGLAP equation can be written in the form
\begin{equation} \label{eq:DGLAP1}
  \frac{\partial\, a(x,k_t^2)}{\partial \log k_t^2}  = \frac{\alpha_S(k_t^2)}{2 \pi}\,\sum_{b}\,\int_x^1\!\dif{z}\,P_{ab }(z)\,b  \left (\frac{x}{z}, k_t^2 \right) - \frac{a(x,k_t^2)}{T_a(k_t^2,\mu^2)}\,\frac{\partial \,T_a (k_t^2,\mu^2)}{\partial \log k_t^2}.
\end{equation}
We define the unintegrated distribution to be
\begin{eqnarray} \label{eq:UPDF}
  f_a(x,k_t^2,\mu^2) &\equiv& \frac{\partial}{\partial \log k_t^2}\left[\,a(x,k_t^2)\,T_a(k_t^2,\mu^2)\,\right]\nonumber \\ &=& T_a(k_t^2,\mu^2)\,\frac{\alpha_S(k_t^2)}{2\pi}\,\sum_{b }\,\int_x^1\!\dif{z}\,P_{ab }(z)\,b \left (\frac{x}{z}, k_t^2 \right).
\end{eqnarray}
This definition is meaningful for $k_t > \mu_0$, where $\mu_0\sim 1$ GeV is the minimum scale for which DGLAP evolution of the conventional parton distributions, $a(x,\mu^2)$, is valid.  Integrating over transverse momentum up to the factorisation scale we find that
\begin{eqnarray}
  \int_{\mu_0^2}^{\mu^2}\!\diff{k_t^2}\,f_a(x,k_t^2,\mu^2) &=& \left[\,a(x,k_t^2)\,T_a(k_t^2,\mu^2)\,\right]_{k_t=\mu_0}^{k_t=\mu} \nonumber \\  &=& a(x,\mu^2) - a(x,\mu_0^2)\,T_a(\mu_0^2,\mu^2).
\end{eqnarray}
Thus, the normalisation condition \eqref{eq:norm} will be exactly satisfied if we define
\begin{equation} \label{eq:smallkt}
  \left.\frac{1}{k_t^2}\,f_a(x,k_t^2,\mu^2)\right\rvert_{k_t<\mu_0} = \frac{1}{\mu_0^2}\,a(x,\mu_0^2)\,T_a(\mu_0^2,\mu^2),
\end{equation}
so that the density of partons in the proton is constant for $k_t<\mu_0$ at fixed $x$ and $\mu$.

So far, we have ignored the singular behaviour of the unregularised splitting kernels, $P_{ab}(z)$, at $z=1$, corresponding to soft gluon emission.  These soft singularities cancel between the real and virtual parts of the DGLAP equation \eqref{eq:DGLAP}.  After resumming the virtual part to all orders in the Sudakov factor \eqref{eq:Sudakov} the singularities must be regulated for the unintegrated distributions to be defined.  The singularities indicate a physical effect that we have not yet accounted for. Here, it is the angular ordering caused by colour coherence, implying a cutoff on the splitting fraction $z$ for those splitting kernels where a real gluon is emitted in the $s$-channel.

We now apply the angular-ordering constraints of Section~\ref{sec:evolution} specifically to the last evolution step. For all other evolution steps, the strong ordering in transverse momentum automatically ensures angular ordering. The condition $z_n\overline{p}_n < \mu$ \eqref{eq:angord} implies
\begin{equation} \label{eq:zcutoff}
  z\frac{k_t}{1-z} < \mu \qquad\iff\qquad z < \frac{\mu}{\mu+k_t},
\end{equation}
where, as before, we have dropped the subscript $n$ specifying the last evolution step. Recall, from the comment below \eqref{eq:a10}, that $\mu$ is entirely determined from the kinematics of the subprocess at the top of the evolution ladder:
\begin{equation} \label{eq:defofmu}
  \mu= Q\frac{x}{\xB}\sqrt{\Xi} =Q\frac{x}{\xB}\sqrt{\frac{1-\beta}{x/\xB-1}}\,.
\end{equation}
Equation~\eqref{eq:zcutoff} applies only to those splitting functions in the real part of the DGLAP equation associated with gluon emission in the $s$-channel.  By unitarity the same form of the cutoff must be chosen in the virtual part. We define $\zeta_{\mathrm{max}}=1-\zeta_{\mathrm{min}}=\mu/(\mu+\kappa_t)$ and insert $\Theta(\zeta_{\mathrm{max}}-\zeta)$ into the Sudakov factor for those splitting functions where a gluon is emitted in the $s$-channel and $\Theta(\zeta-\zeta_{\mathrm{min}})$ where a gluon is emitted in the $t$-channel.  Note that there is no ``coherence'' effect for quark (fermion) emission and therefore the phase space available for quark emission is not restricted by the angular-ordering condition~\eqref{eq:zcutoff}.\footnote{This is in contrast to Ref.~\cite{Kimber:2001sc}, where a cutoff on the splitting fraction was applied both to quark and gluon emissions.  Also, in \cite{Kimber:2001sc}, the scale $\mu$ was treated as a free parameter, which was chosen to be the hard scale of the subprocess, or a combination of hard scales.  Here we fix $\mu$ using \eqref{eq:defofmu}.}

The precise expressions for the unintegrated quark and gluon distributions are
\begin{equation} \label{eq:a11}
  f_q(x,k_t^2,\mu^2) = T_q(k_t^2,\mu^2)\,\frac{\alpha_S(k_t^2)}{2\pi}\,\int_x^1\!\dif{z}\;\left[\,P_{qq}(z)\,\frac{x}{z}q\left(\frac{x}{z},k_t^2\right)\,\Theta\left(\frac{\mu}{\mu+k_t}-z\right)+ P_{qg}(z)\,\frac{x}{z}g\left(\frac{x}{z},k_t^2\right)\,\right]
\end{equation}
and
\begin{equation}\label{eq:a12}
  f_g(x,k_t^2,\mu^2) = T_g(k_t^2,\mu^2)\,\frac{\alpha_S(k_t^2)}{2\pi}\,\int_x^1\!\dif{z}\;\left[\sum_q P_{gq}(z)\,\frac{x}{z}q\left(\frac{x}{z},k_t^2\right) + P_{gg}(z)\,\frac{x}{z}g\left(\frac{x}{z},k_t^2\right)\,\Theta\left(\frac{\mu}{\mu+k_t}-z\right)\,\right].
\end{equation}

The exponent of the quark Sudakov factor can be simplified using the fact that $P_{gq}(1-\zeta )=P_{qq}(\zeta )$.  Then
\begin{equation}
  \int_0^{\zeta_{{\rm max}}}\!\dif{\zeta }\;\zeta \,P_{qq}(\zeta ) + \int_{\zeta_{{\rm min}}}^1\!\dif{\zeta}\;\zeta \,P_{gq}(\zeta ) = \frac{1}{2}\left[\int_0^{\zeta_{{\rm max}}}\!\dif{\zeta }\,P_{qq}(\zeta)+\int_{\zeta_{{\rm min}}}^1\!\dif{\zeta }\,P_{gq}(\zeta) \right] = \int_0^{\zeta_{{\rm max}}}\!\dif{\zeta}\,P_{qq}(\zeta),
\end{equation}
so that
\begin{equation}
  T_q(k_t^2,\mu^2) = \exp\left(-\int_{k_t^2}^{\mu^2}\!\diff{\kappa_t^2}\,\frac{\alpha_S(\kappa_t^2)}{2\pi}\,\int_0^{\zeta_{{\rm max}}}\!\dif{\zeta }\,P_{qq}(\zeta )\right).
\end{equation}
Similarly, the exponent of the gluon Sudakov factor can be simplified by exploiting the symmetry $P_{qg}(1-\zeta)=P_{qg}(\zeta)$. We have
\begin{equation}
  \sum_{q}\int_0^{1}\!\dif{\zeta }\;\zeta \,P_{qg}(\zeta ) = 2n_F\int_0^{1}\!\dif{\zeta}\;\frac{1}{2}\,P_{qg}(\zeta ) = n_F\,\int_0^1\!\dif{\zeta }\,P_{qg}(\zeta ),
\end{equation}
so that the gluon Sudakov factor is
\begin{equation}
  T_g(k_t^2,\mu^2) = \exp\left(-\int_{k_t^2}^{\mu^2}\!\diff{\kappa_t^2}\,\frac{\alpha_S(\kappa_t^2)}{2\pi}\,\left( \int_{\zeta_{{\rm min}}}^{\zeta_{{\rm max}}}\!\dif{\zeta }\;\zeta \,P_{gg}(\zeta ) + n_F\,\int_0^1\!\dif{\zeta}\,P_{qg}(\zeta)\right)\right),
\end{equation}
where $n_F$ is the active number of quark--antiquark flavours into which the gluon may split.

It is important to note that the starting point of our derivation is the LO DGLAP equation \eqref{eq:DGLAP}, with LO DGLAP splitting kernels and one-loop running coupling.  Therefore, in order for the normalisation \eqref{eq:norm} to be satisfied, it is essential that we use a LO parton set where the integrated parton distributions have been determined using the same splitting kernels and running coupling.  In Ref.~\cite{Kimber:2001sc}, the MRST99 parton set \cite{Martin:1999ww} was used, which has been determined using NLO DGLAP splitting kernels and two-loop running coupling, therefore \eqref{eq:norm} was found not to be satisfied.  Also, in Ref.~\cite{Kimber:2001sc} the angular-ordering constraints were not correctly applied and the Sudakov factor $T_a(\mu_0^2,\mu^2)$ was omitted from \eqref{eq:smallkt}.  We have checked numerically that our refined prescription now gives the exact normalisation of \eqref{eq:norm}.

\section{Calculating the cross section} \label{sec:newfact}

We have defined unintegrated parton distributions, $f_a(x,k_t^2,\mu^2)$, valid for all values of $x$ for both the quark and gluon. This was done by assuming that the transverse momentum of the parton is generated entirely in the last evolution step and then imposing constraints from angular ordering to regulate the soft gluon singularities.  It now remains to specify the prescription for calculating observables such as cross sections.

The penultimate parton in the evolution chain has 4-momentum $k_{n-1}=xp/z$.  In the final evolution step, it splits into a parton with 4-momentum $k_n\equiv k=x\,p-\beta\,q^\prime+k_\perp$ and an emitted parton of 4-momentum $p_n=k_{n-1}-k_n$.  The variable $\beta$ is specified by the on-shell condition, $p_n^2=0$, which gives
\begin{equation}  \label{eq:defbeta}
  \beta = \frac{\xB}{x}\,\frac{z}{(1-z)}\,\frac{k_t^2}{Q^2}\,.
\end{equation}
Hence $k^2 = -k_t^2/(1-z)$.  The rapidity of the emitted parton is
\begin{equation} \label{eq:etaBj}
  \eta^{{\rm Breit}} = \frac{1}{2}\log\frac{p_n^+}{p_n^-} = \frac{1}{2}\log\frac{x\,(1-z)}{\xB \,z\,\beta}.
\end{equation}

\begin{figure}
  \begin{center}
    \includegraphics[width=0.35\textwidth]{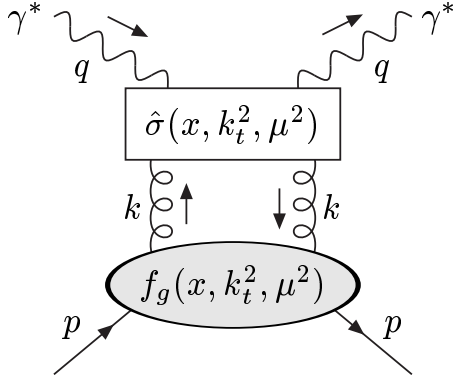}
    \caption{Schematic picture of the $k_t$-factorisation formula \eqref{eq:ktfactCCFM} \label{fig:ktfact}}
  \end{center}
\end{figure}

In the small $x$ regime, where gluons dominate, the main contribution comes from the $z\to0$ limit, where $k\simeq x\,p+k_\perp$, $k^2\simeq -k_t^2$ and the emitted gluon has a large positive rapidity.  In this case, observables can be calculated from the $k_t$-factorisation prescription.  For example, for deep-inelastic scattering, the cross sections for the scattering of a virtual photon with transverse or longitudinal polarisation can be expressed in the form
\begin{equation} \label{eq:ktfactCCFM}
  \sigma_{T,L}^{\gamma^*p} = \int_0^1\!\diff{x}\,\int_0^{\infty}\!\diff{k_t^2}\;f_g(x,k_t^2,\mu^2)\;\hat{\sigma}_{T,L}^{\gamma^*{g}}(x,k_t^2,\mu^2),
\end{equation}
see Fig.~\ref{fig:ktfact}.
At small $x$, we would expect that the leading $\log(1/x)$ terms would need to be resummed. However, in Ref.~\cite{Kimber:2001sc} it was found that the unintegrated gluon based on a unified BFKL-DGLAP equation was very similar to the unintegrated gluon calculated purely from the DGLAP equation, as in Section \ref{sec:defunint}.

In \cite{Kimber:2001sc} the $k_t$-factorisation approach was used to calculate the unintegrated gluon contribution to the proton structure function $F_2(\xB,Q^2)$.  The unintegrated quark contribution was estimated in a rather \emph{ad hoc} manner.  In \cite{Kimber:1999xc} the normal on-shell partonic cross section was evaluated with off-shell kinematics to estimate the cross section for prompt photon hadroproduction.  Again, the $z$-dependence of the hard-scattering coefficient was neglected.

\subsection{Generalising $k_t$-factorisation}

Clearly, it is desirable to formulate a more general prescription for the calculation of cross sections using unintegrated parton distributions.  This prescription should be valid for both quarks and gluons and without taking the limit $z\to0$.  The ``partonic cross section'' will necessarily have some $z$-dependence, therefore we must consider parton distributions, $f_a(x,z,k_t^2,\mu^2)$, \emph{doubly}-unintegrated over both $z$ and $k_t^2$, satisfying the normalisation conditions
\begin{equation} \label{eq:znorm0}
  \int_x^1\!\dif{z}\,f_a(x,z,k_t^2,\mu^2)=f_a(x,k_t^2,\mu^2)
\end{equation}
and
\begin{equation} \label{eq:znorm}
  \int_x^1\!\dif{z}\int_0^{\mu^2}\!\diff{k_t^2}\,f_a(x,z,k_t^2,\mu^2)=a(x,\mu^2).
\end{equation}
These normalisation conditions are only satisfied for fixed $x$ and $\mu$, independent of the integration variables $z$ or $k_t$.  Apart from the angular-ordering constraints, the distributions may be obtained from \eqref{eq:UPDF}:
\begin{equation} \label{eq:UPDFz}
  f_a(x,z,k_t^2,\mu^2) = T_a(k_t^2,\mu^2)\,\frac{\alpha_S(k_t^2)}{2\pi}\sum_{b }P_{ab }(z)\,b\left(\frac{x}{z},k_t^2\right)\,.
\end{equation}
The explicit forms, including the constraints, follow from \eqref{eq:a11} and \eqref{eq:a12}:
\begin{equation}
  f_q(x,z,k_t^2,\mu^2) = T_q(k_t^2,\mu^2)\,\frac{\alpha_S(k_t^2)}{2\pi}\;\left[\,P_{qq}(z)\,\frac{x}{z}q\left(\frac{x}{z},k_t^2\right)\,\Theta\left(\frac{\mu}{\mu+k_t}-z\right) + P_{qg}(z)\,\frac{x}{z}g\left(\frac{x}{z},k_t^2\right)\,\right]
\end{equation}
and
\begin{equation}
  f_g(x,z,k_t^2,\mu^2) = T_g(k_t^2,\mu^2)\,\frac{\alpha_S(k_t^2)}{2\pi}\;\left[\sum_q P_{gq}(z)\,\frac{x}{z}q\left(\frac{x}{z},k_t^2\right) + P_{gg}(z)\,\frac{x}{z}g\left(\frac{x}{z},k_t^2\right)\,\Theta\left(\frac{\mu}{\mu+k_t}-z\right)\,\right].
\end{equation}

The universal factorisation formula involving these doubly-unintegrated distributions, analogous to \eqref{eq:ktfactCCFM}, is
\begin{equation} \label{eq:newktfact}
  \sigma_{T,L}^{\gamma^*p} = \sum_a \int_0^1\!\diff{x}\,\int_x^1\!\dif{z}\,\int_0^{\infty}\!\diff{k_t^2}\;f_a(x,z,k_t^2,\mu^2)\;\hat{\sigma}_{T,L}^{\gamma^*a}(x,z,k_t^2,\mu^2),
\end{equation}
where $\hat{\sigma}_{T,L}^{\gamma^*a}$ are now the partonic cross sections for an incoming parton with (plus) momentum fraction $x$ and transverse momentum $k_t$, which has split from a parent parton with (plus) momentum fraction $x/z$ and zero transverse momentum.  We will refer to this generalised form of $k_t$-factorisation as $(z,k_t$)-factorisation.

There will be an effective upper bound on the $k_t$ integration from kinematics, but note that there is no restriction to the domain $k_t<\mu$, as in conventional DGLAP calculations.  For $k_t>\mu$, the Sudakov form factors $T_a(k_t^2,\mu^2)$ are defined to be 1.

Taking the limit $z\to0$ of $\hat{\sigma}_{T,L}^{\gamma^*g}(x,z,k_t^2,\mu^2)$ in \eqref{eq:newktfact} we essentially recover the conventional $k_t$-factorisation prescription of \eqref{eq:ktfactCCFM}. Alternatively, in the limit $k_t\ll Q$, we  recover the conventional collinear factorisation prescription.

Note that $f_a(x,z,k_t^2,\mu^2)$ is undefined for $k_t<\mu_0\sim 1$ GeV and also that \eqref{eq:smallkt} no longer applies since there is now a $z$-dependence involved.  To approximate the $k_t<\mu_0$ contribution of \eqref{eq:newktfact}, we choose to take the collinear limit $k_t\ll Q$ in the hard-scattering coefficients, so that
\begin{equation}
  \left.\hat{\sigma}_{T,L}^{\gamma^*a}(x,z,k_t^2,\mu^2)\right\rvert_{k_t<\mu_0} = \lim_{k_t\ll Q}\hat{\sigma}_{T,L}^{\gamma^*a}(x,z,k_t^2,\mu^2) \equiv \hat{\sigma}_{T,L}^{\gamma^*a}(x,\mu^2).
\end{equation}
We then make the replacement
\begin{equation}
  \int_x^1\!\dif{z}\,\int_0^{\mu_0^2}\!\diff{k_t^2}\,f_a(x,z,k_t^2,\mu^2) = a(x,\mu_0^2)\,T_a(\mu_0^2,\mu^2),
\end{equation}
so that the $(z,k_t)$-factorisation formula \eqref{eq:newktfact} becomes
\begin{equation} \label{eq:newktfact1}
  \sigma_{T,L}^{\gamma^*p} = \sum_a \int_0^1\!\diff{x}\,\left[a(x,\mu_0^2)\,T_a(\mu_0^2,\mu^2)\,\hat{\sigma}_{T,L}^{\gamma^*a}(x,\mu^2) + \int_x^1\!\dif{z}\,\int_{\mu_0^2}^{\infty}\!\diff{k_t^2}\;f_a(x,z,k_t^2,\mu^2)\;\hat{\sigma}_{T,L}^{\gamma^*a}(x,z,k_t^2,\mu^2)\right].
\end{equation}
In the first term, the limit $k_t\ll Q$ must also be taken in the expressions determining $x$ and $\mu$.  In the following, we will use \eqref{eq:newktfact} for brevity, with the understanding that the $k_t<\mu_0$ region is to be dealt with as in \eqref{eq:newktfact1}.

\subsection{Motivation for the ($z,k_t$)-factorisation formula}

\begin{figure}
  \begin{center}
    \includegraphics[width=1.0\textwidth]{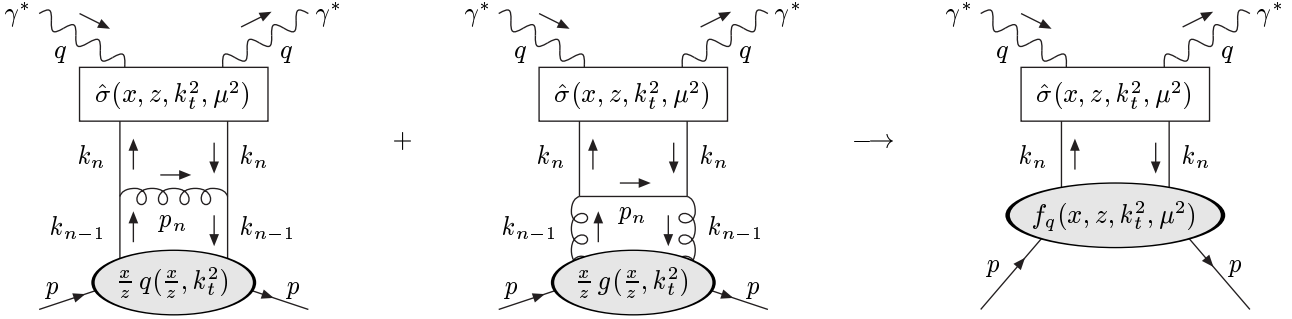}
    \caption{Verification of $(z,k_t)$-factorisation for the doubly-unintegrated quark distribution, $f_q(x,z,k_t^2,\mu^2)$, shown in the final diagram. In the first two diagrams the penultimate parton in the DGLAP evolution chain, with 4-momentum $k_{n-1}=xp/z$, splits into a quark with 4-momentum $k_n\equiv k=x\,p-\beta\,q^\prime+k_\perp$. \label{fig:zktfactq}}
  \end{center}
\end{figure}

\begin{figure}
  \begin{center}
    \includegraphics[width=1.0\textwidth]{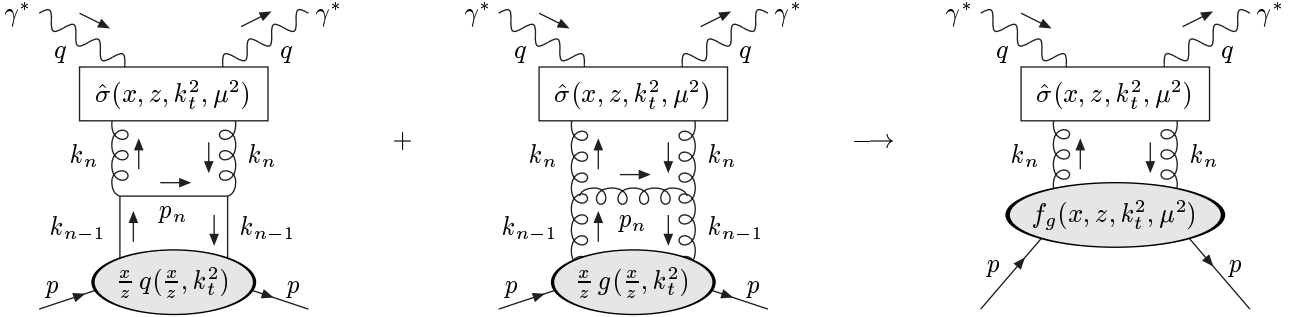}
    \caption{Verification of $(z,k_t)$-factorisation for the doubly-unintegrated gluon distribution, $f_g(x,z,k_t^2,\mu^2)$, shown in the final diagram. In the first two diagrams the penultimate parton in the DGLAP evolution chain, with 4-momentum $k_{n-1}=xp/z$, splits into a gluon with 4-momentum $k_n\equiv k=x\,p-\beta\,q^\prime+k_\perp$. \label{fig:zktfactg}}
  \end{center}
\end{figure}

At this stage, it is perhaps unclear exactly how we should calculate the partonic cross sections, $\hat{\sigma}_{T,L}^{\gamma^*a}(x,z,k_t^2,\mu^2)$, since the incoming parton is now off-shell with virtuality $k^2=-k_t^2/(1-z)$, and so the usual $k_t$-factorisation approach does not apply. This issue can be clarified by starting with the collinear factorisation formula one rung down. That is,
\begin{equation} \label{eq:collfactb}
  \sigma_{T,L}^{\gamma^*p} = \sum_{b} \int_0^1\!\diff{(x/z)}\;b(x/z,k_t^2)\;\hat{\sigma}_{T,L}^{\gamma^*b}(x/z,k_t^2),
\end{equation}
where we have chosen the factorisation scale to be $k_t$, and $b$ is the penultimate parton in the evolution chain of Fig.~\ref{fig:evchain}, so that $\hat\sigma^{\gamma^*b}$ incorporates the last evolution step.  From Fig.~\ref{fig:evchain} we see that the parton $b$, with 4-momentum $k_{n-1}=xp/z$, splits into a parton of type $a$ with 4-momentum $k_n\equiv k=x\,p-\beta\,q^\prime+k_\perp$, which then goes on to initiate the hard subprocess at a scale $\mu$ given by \eqref{eq:defofmu}. To derive formula~\eqref{eq:newktfact} we need to show that the partonic cross section $\hat{\sigma}^{\gamma^*b}$ can be factorised to give a partonic cross section for the $\gamma^*a$ subprocess, $\hat{\sigma}^{\gamma^*a}$, with the remainder being absorbed into the definition of the doubly-unintegrated density, $f_a(x,z,k_t^2,\mu^2)$. This idea is illustrated in Fig.~\ref{fig:zktfactq} for the doubly-unintegrated quark distribution, and in Fig.~\ref{fig:zktfactg} for the doubly-unintegrated gluon distribution.

The squared matrix element can be factorised if we assume the LLA, so that only the leading $1/k_t^2$ term is kept and terms not giving a logarithmic divergence in the collinear limit are neglected.  We find that
\begin{equation} \label{eq:m2fact}
  \lvert\mathcal{M}_{T,L}^{\gamma^*b}\rvert^2 = 16\pi^2\frac{(1-z)}{zk_t^2}\frac{\alpha_S(k_t^2)}{2\pi} \sum_a P_{ab}(z)\ \lvert\mathcal{M}_{T,L}^{\gamma^*a}\rvert^2\ \times \ \left[1+\mathcal{O}(\beta)\right],
\end{equation}
where $\lvert\mathcal{M}^{\gamma^*a}\rvert^2$ represents the squared matrix element of the $\gamma^*a$ subprocess, containing one power of $\alpha_S$ less than $\lvert\mathcal{M}^{\gamma^* b}\rvert^2$.  We have used this method to derive the form of all four splitting kernels, $P_{ab}(z)$.  It is crucial that we adopt a physical gauge for the gluon so that the splitting kernels are obtained from only the ladder-type diagrams.

The extra terms of \eqref{eq:m2fact} are proportional to $\beta$ and so are negligible for either $k_t\ll Q$ or $z\to 0$.  Away from these limits, it is far from obvious that these ``beyond LLA''  terms will be small, a necessary condition for the factorisation to hold.  We will observe that the main effect of the extra terms is to suppress the contribution from large $z$ for gluon emission.  In our approach, we achieve the same effect with angular ordering, so the extra terms may be neglected.

$\lvert\mathcal{M}^{\gamma^* a}\rvert^2$ must also be evaluated in the LLA for the factorisation to hold, so terms of $\mathcal{O}(k_t^2/Q^2)$ should be neglected when calculating this. This amounts to the replacement $k\to x\,p$ in the numerator of $\lvert\mathcal{M}^{\gamma^* a}\rvert^2$, but not in the propagator virtualities in the denominator.   Of course, $x$ may have some $k_t$ dependence from kinematics, so some terms beyond the LLA are included in this respect.

The phase space $\dif\Phi^{\gamma^* b}$ can be factorised easily to give the phase space $\dif\Phi^{\gamma^* a}$:
\begin{eqnarray}
  \dif\Phi^{\gamma^* b} &=& \dif\Phi^{\gamma^* a}\quad\frac{\dif^3\vec{p_n}}{2p_n^0\,(2\pi)^3} \nonumber \\
  &=& \dif\Phi^{\gamma^* a}\quad\frac{1}{(2\pi)^3}\;\dif^4k\;\delta(\,p_n^2\,) \label{eq:psfact} \\
  &=& \dif\Phi^{\gamma^* a}\quad\frac{1}{16\pi^2}\;\dif{x}\,\dif{k_t^2}\,\frac{z}{x(1-z)}, \nonumber
\end{eqnarray}
where we have used $\dif^4k=p\cdot q\;\dif{x}\,\dif\beta\,\dif^2\vec{k_t}$ and $\dif^2\vec{k_t}=k_t\,\dif{k_t}\,\dif{\phi}=\pi\dif{k_t^2}$, after integrating over the azimuthal angle $\phi$.  The $\beta$ integration absorbs the delta function, determining $\beta$ as given by \eqref{eq:defbeta}.

The partonic flux factor $F^{\gamma^* a}$ is not well defined since the parton $a$ is off-shell and non-collinear with the photon.  As in conventional $k_t$-factorisation, we define it to be\footnote{Choosing another definition for the flux factor is a NLL effect.}
\begin{equation}
  F^{\gamma^* a} \equiv z\,F^{\gamma^* b} = z\;4\,k_{n-1}\cdot q = 4x\,p\cdot q.
\end{equation}
Finally, we have the relationship
\begin{equation} \label{eq:dsighat}
  \dif{\hat{\sigma}_{T,L}^{\gamma^*{b}}} = \dif\Phi^{\gamma^*{b}}\,\lvert\mathcal{M}_{T,L}^{\gamma^*{b}}\rvert^2 \,/\, F^{\gamma^*{b}} = \diff{x}\,\diff{k_t^2}\,z\frac{\alpha_S(k_t^2)}{2\pi}\,\sum_a P_{ab}(z)\,\dif\hat{\sigma}_{T,L}^{\gamma^*a}.
\end{equation}
To calculate the hadronic cross section, we insert \eqref{eq:dsighat} into \eqref{eq:collfactb}
\begin{eqnarray}
  \dif{\sigma_{T,L}^{\gamma^*p}} &=& \sum_{b} \diff{(x/z)}\;b(x/z,k_t^2)\;\dif{\hat{\sigma}_{T,L}^{\gamma^*{b}}} \nonumber \\
  &=& \sum_{b} \diff{z}\;\diff{x}\;\diff{k_t^2}\;z\frac{\alpha_S(k_t^2)}{2\pi}\,\sum_a P_{ab}(z)\;b(x/z,k_t^2)\;\dif{\hat{\sigma}_{T,L}^{\gamma^*a}} \\
  &\to& \sum_a \diff{x}\;\dif{z}\;\diff{k_t^2}\;f_a(x,z,k_t^2,\mu^2)\;\dif\hat{\sigma}_{T,L}^{\gamma^*a}(x,z,k_t^2,\mu^2), \nonumber
\end{eqnarray}
where in the last step we recognise the ``real'' part of the doubly-unintegrated distribution given in \eqref{eq:UPDFz}.  The $(z,k_t)$-factorisation formula \eqref{eq:newktfact} follows easily.

\section{Application to inclusive jet production in DIS} \label{sec:jetprod}

The simplest process that we can consider to illustrate the use of the doubly-unintegrated parton distributions is current jet production in DIS. The subprocess is simply $\gamma^*q\to q$ at the top of the evolution chain. In the normal collinear factorisation approach, this diagram gives the parton model prediction for the structure function $F_2(\xB,Q^2)$. Indeed, measurements of $F_2(\xB,Q^2)$ are used to determine the integrated quark distribution $q(x,\mu^2)$. In the new $(z,k_t)$-factorisation framework of Section \ref{sec:newfact}, where the incoming quark has transverse momentum $\vec{k_t}$, we produce a current jet with transverse momentum $\vec{k_t}$ and transverse energy $E_T=k_t$.  The parton emitted in the last evolution step will emerge with transverse momentum $-\vec{k_t}$ and transverse energy $E_T=k_t$.

The inclusive jet cross section counts all jets passing the required cuts. The cross section, integrated over bins in $y$, $Q^2$, $E_T$ and $\eta$ is
\begin{eqnarray} \label{eq:jetcross}
  \sigma({\rm jet}) & = & \int_{y_{\rm min}}^{y_{\rm max}}\!\!\!\dif y\,\int_{Q^2_{\rm min}}^{Q^2_{\rm max}}\!\!\!\dif Q^2\;\frac{\alpha}{2\pi yQ^2}\left[(1+(1-y)^2)\,\sigma_T^{\gamma^*p} + 2(1-y)\,\sigma_L^{\gamma^*p}\right] \; \\ &&\qquad\times\;\sum_{\mathrm{jets}}\;\Theta\left(E_T-{E_T}_{\rm min}\right)\,\Theta\left({E_T}_{\rm max}-E_T\right)\;\Theta\left(\eta-\eta_{\rm min}\right)\,\Theta\left(\eta_{\rm max}-\eta\right), \nonumber 
\end{eqnarray}
where $y=Q^2/(\xB s)$, and where the sum is over all jets with transverse energy $E_T$ and rapidity $\eta$.  The differential cross sections are easily obtained by dividing by the size of the bin, for example,
\begin{equation}
  \frac{\dif\sigma}{\dif E_T} = \sigma({\rm jet})/({E_T}_{\rm max}-{E_T}_{\rm min}).
\end{equation}

In Section \ref{sec:newfact} we gave the general prescription for calculating the cross section.  Recall that it was necessary to consider the doubly-unintegrated parton distributions, $f_a(x,z,k_t^2,\mu^2)$, to keep the precise kinematics in the subprocess, without taking the limit $z\to 0$.  We now check that this prescription reproduces with good accuracy the conventional LO QCD calculation with integrated partons, where all $\mathcal{O}(\alpha_S)$ diagrams are included, not just the ones which give the leading $\dif{k_t^2}/k_t^2$ term.  With the $(z,k_t)$-factorisation approach, in addition to the jets produced in the hard subprocess, we must also count the parton emitted in the last evolution step with transverse energy $E_T=k_t$ and rapidity given by \eqref{eq:etaBj}.

We also explain how the prescription may be extended to higher orders in perturbation theory.  The conventional NLO QCD diagrams are at $\mathcal{O}(\alpha_S^2)$.  These include all real and virtual $\mathcal{O}(\alpha_S)$ corrections to the LO QCD diagrams.  The hard-scattering coefficients obtained from these diagrams are convoluted with NLO integrated partons, $a(x,\mu^2)$, satisfying the DGLAP equation with two-loop $\alpha_S$ and splitting kernels.  Several codes are available which include these NLO QCD calculations.  There is no longer a one-to-one correspondence between partons and jets.  The 4-momenta of the outgoing partons should be passed through a jet algorithm to assign the partons to jets.  At NLO in the $(z,k_t)$-factorisation approach, we continue to use the LO doubly-unintegrated partons constructed in Section \ref{sec:defunint} and only calculate the $\mathcal{O}(\alpha_S)$ diagrams expected to dominate.

\subsection{Collinear factorisation approach at LO} \label{sec:LOQCD}

In the collinear approximation, the LO QCD Feynman diagrams are at $\mathcal{O}(\alpha_S)$.  These are the boson-gluon fusion process, $\gamma^* g \to q\bar{q}$, and the QCD Compton process, $\gamma^* q \to qg$, illustrated in Fig.~\ref{fig:LOjets}(b).  These partonic processes give rise to two jets with equal transverse energy and opposite transverse momentum.  There is a one-to-one correspondence between partons and jets.  There are no singularities to be regulated and no cutoff is imposed on gluon emission.

\begin{figure}
  \begin{center}
    \includegraphics[width=0.8\textwidth]{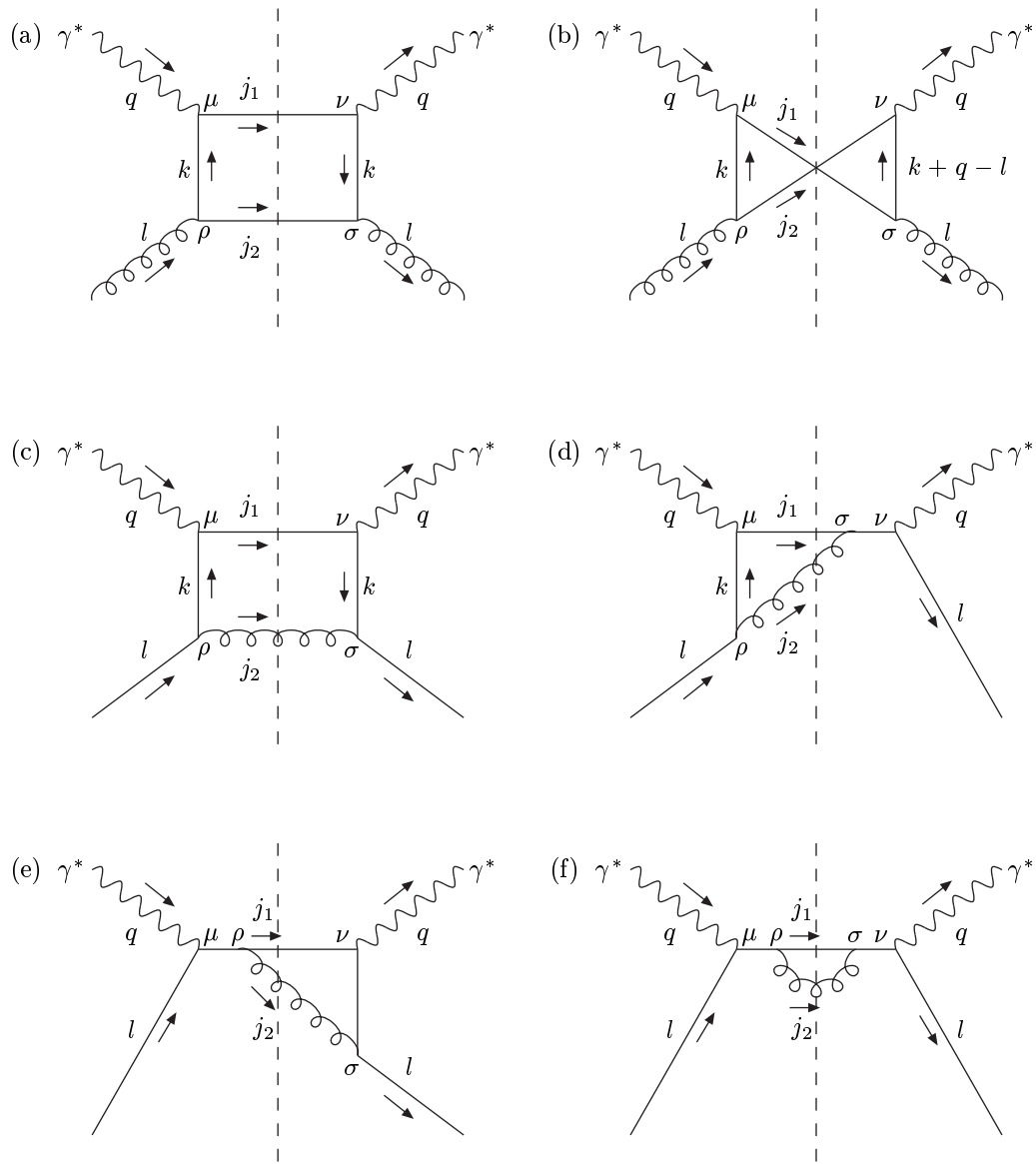}
    \caption{Cut diagrams contributing to inclusive jet production in LO QCD.\label{fig:LOQCDcut}}
  \end{center}
\end{figure}

We now explain a few of the details involved since this calculation offers valuable insights into the $(z,k_t)$-factorisation approach.    The cut diagrams are illustrated in Fig.~\ref{fig:LOQCDcut}.  Note that the direction of fermion number flow is not indicated in these diagrams.  The arrows indicate only the direction of the labelled 4-momentum and this is taken to be the same for both quarks and antiquarks.  The contribution from diagrams (a) to (f) to $\sigma^{\gamma^* p}$ need to be added together.  Diagrams (a) to (d) have the same kinematics, so we calculate them first.  We label the 4-momenta by
\begin{gather}
  q=q^\prime-\xB\,p,\qquad l=\frac{x}{z}\,p,\qquad k=x\,p-\beta\,q^\prime+k_\perp,\\
  j_1=k+q=(x-\xB)\,p+(1-\beta)\,q^\prime+k_\perp\qquad{\rm and}\qquad j_2=l-k=\frac{x}{z}(1-z)\,p+\beta\,q^\prime-k_\perp,\notag
\end{gather}
with $x\ge \xB$.  The 2-body phase space is
\begin{equation}
  \dif \Phi^{\gamma^*a} = (2\pi)^4\,\delta^{(4)}(l+q-j_1-j_2)\,\frac{\dif^4 j_1}{(2\pi)^3}\,\delta(j_1^2)\,\frac{\dif^4 j_2}{(2\pi)^3}\,\delta(j_2^2) \ =\  \frac{\dif^4 k}{4\pi^2}\,\delta(j_1^2)\,\delta(j_2^2).
\end{equation}
The two delta functions can be used to determine $\beta$ and $x$:
\begin{equation} \label{eq:x_pm}
  \beta = \frac{\xB\,z\,r}{x\,(1-z)}\quad{\rm and}\quad x_\pm =
  \frac{\xB}{2(1-z)}\left(1-z+r\pm\sqrt{(1-z+r)^2-4rz(1-z)}\right),
\end{equation}
where $r\equiv k_t^2/Q^2$.  The flux factor is $F^{\gamma^*a}=4\,l\cdot q$, so that
\begin{equation}
  \frac{\dif \Phi^{\gamma^*a}}{F^{\gamma^*a}} = \frac{\dif k_t^2}{16\pi}\left(\frac{\xB}{Q^2}\right)^2\sum_{x=x_\pm}\frac{z^2}{x^2(1-z)}\frac{1}{1-\xB\beta/x}.
\end{equation}
In practice, the condition $x\ge \xB$ ensures that only the $x=x_+$ solution contributes.

The squared matrix elements of all six diagrams can be written in the form
\begin{equation} \label{eq:Msquared}
  \lvert \mathcal{M}_{T,L}^{\gamma^*a} \rvert^2 = \frac{1}{2}\,e^2\,g^2\;M^{\mu\nu}\;\epsilon_\mu(q,\lambda)\epsilon_\nu^*(q,\lambda),
\end{equation}
where $\lambda$ is either $T$ or $L$ and the initial factor of $1/2$ is to average over the helicity of the incoming parton.  Appropriate scales have been chosen for the two running couplings, $e^2=4\pi\alpha(Q^2)$ and $g^2=4\pi\alpha_S(k_t^2)$.  We have
\begin{eqnarray}
  {\rm (a)}\qquad M^{\mu\nu} &=& \left(\sum_q e_q^2\right) T_R\frac{1}{k^4}{\rm Tr}\left[\slashed{k}\gamma^\rho\slashed{j_2}\gamma^\sigma\slashed{k}\gamma^\nu\slashed{j_1}\gamma^\mu\right]d_{\rho\sigma}(l), \\
  {\rm (b)}\qquad M^{\mu\nu} &=& \left(\sum_q e_q^2\right) T_R\frac{1}{k^2}\frac{1}{(k+q-l)^2}{\rm Tr}\left[\slashed{k}\gamma^\rho\slashed{j_2}\gamma^\nu(\slashed{k}+\slashed{q}-\slashed{l})\gamma^\sigma\slashed{j_1}\gamma^\mu\right]d_{\rho\sigma}(l), \\
  {\rm (c)}\qquad M^{\mu\nu} &=& e_q^2 C_F\frac{1}{k^4}{\rm Tr}\left[\slashed{k}\gamma^\rho\slashed{l}\gamma^\sigma\slashed{k}\gamma^\nu\slashed{j_1}\gamma^\mu\right]d_{\rho\sigma}(j_2), \\
  {\rm (d)}\qquad M^{\mu\nu} &=& e_q^2 C_F\frac{1}{k^2}\frac{1}{(l+q)^2}{\rm Tr}\left[\slashed{k}\gamma^\rho\slashed{l}\gamma^\nu(\slashed{l}+\slashed{q})\gamma^\sigma\slashed{j_1}\gamma^\mu\right]d_{\rho\sigma}(j_2),
\end{eqnarray}
where the colour factors are $T_R=1/2$ and $C_F=4/3$.  In an axial gluon gauge with a light-like gauge fixing vector $n=q^\prime$,
\begin{equation}
  d_{\rho\sigma}(l) \equiv -g_{\rho\sigma} + \frac{n_\rho l_\sigma + l_\rho n_\sigma}{n\cdot l}\,.
\end{equation}

For diagrams (e) and (f) of Fig.~\ref{fig:LOQCDcut}, the 4-momenta can be parameterised as
\begin{gather}
  q=q^\prime-\xB\,p,\qquad l=X\,p,\qquad j_1=\xi\,p+b\,q^\prime+k_\perp\\
  {\rm and}\qquad j_2=l+q-j_1=(X-\xB-\xi)\,p+(1-b)\,q^\prime-k_\perp.\notag
\end{gather}
with $0\le\xi \le X-\xB\le 1$ and $0\le b\le 1$.  This time the 2-body phase space determines
\begin{equation}
  b=\xB r/\xi \qquad {\rm and}\qquad \xi_\pm = \frac{1}{2}\left\{X-\xB\pm\sqrt{(X-\xB)(X-(1+4r)\xB)} \right\}.
\end{equation}
Dividing the phase space by the flux factor gives
\begin{equation}
  \frac{\dif \Phi^{\gamma^*q}}{F^{\gamma^*q}} = \frac{\dif k_t^2}{16\pi}\,\left(\frac{\xB}{Q^2}\right)^2\,\sum_{\xi=\xi_\pm}\frac{1}{X\xi}\,\frac{1}{\lvert 1-b(X-\xB)/\xi\rvert}\,,
\end{equation}
and the squared matrix elements are
\begin{eqnarray}
  {\rm (e)}\qquad M^{\mu\nu} &=& e_q^2 C_F\frac{1}{(l+q)^2}\frac{1}{(j_1-q)^2}{\rm Tr}\left[\slashed{l}\gamma^\sigma(\slashed{j_1}-\slashed{q})\gamma^\nu\slashed{j_1}\gamma^\rho(\slashed{l}+\slashed{q})\gamma^\mu\right]d_{\rho\sigma}(j_2), \\
  {\rm (f)}\qquad M^{\mu\nu} &=& e_q^2 C_F\frac{1}{(l+q)^2}\frac{1}{(l+q)^2}{\rm Tr}\left[\slashed{l}\gamma^\nu(\slashed{l}+\slashed{q})\gamma^\sigma\slashed{j_1}\gamma^\rho(\slashed{l}+\slashed{q})\gamma^\mu\right]d_{\rho\sigma}(j_2).
\end{eqnarray}

Averaging over the transverse photon polarisations in \eqref{eq:Msquared}, we have
\begin{equation} \label{eq:sumepsT}
  \epsilon_\mu(q,T)\,\epsilon_\nu^*(q,T) \to -\frac{1}{2}g_{\mu\nu}^\perp,
\end{equation}
while demanding that the longitudinal polarisation vector is normalised, $[\epsilon(q,L)]^2=1$, and satisfies the Lorentz condition, $q\cdot\epsilon(q,L) = 0$, leads to
\begin{equation} \label{eq:epsL}
  \epsilon_\mu(q,L)= \frac{1}{Q}(2\xB p_\mu + q_\mu).
\end{equation}
Gauge invariance ensures that the $q_\mu$ term does not contribute to the squared matrix element if all diagrams are included, courtesy of the Ward identity:
\begin{equation} \label{eq:WardID}
  q_\mu M^{\mu\nu} = 0 = q_\nu M^{\mu\nu}.
\end{equation}
Therefore, we are free to neglect the $q_\mu$ term of \eqref{eq:epsL} from the outset, so that
\begin{equation} \label{eq:sumepsL}
  \epsilon_\mu(q,L)\,\epsilon_\nu^*(q,L) \to \frac{4\xB^2}{Q^2}p_\mu p_\nu.
\end{equation}
Finally, the contribution to the $\gamma^*p$ cross section from Fig.~\ref{fig:LOQCDcut} (a), (b), (c) and (d) is
\begin{eqnarray} \label{eq:LOQCDabcd}
  \sigma^{\gamma^*p}_{T,L} & = & \sum_q \frac{4\pi^2\alpha e_q^2}{Q^2}\,\int_x^1\!\dif{z}\,\int_0^\infty\!\diff{k_t^2}\;\sum_{x=x_\pm}\frac{\xB/x}{1-\xB\beta/x}\,\frac{\alpha_S(k_t^2)}{2\pi}\\
  && \qquad\qquad \times\left\{P_{qg}(z)\,\frac{x}{z}g(\frac{x}{z},\mu^2)\left[\mathcal{C}_{T,L}^{a}+\mathcal{C}_{T,L}^{b}\right]+P_{qq}(z)\,\frac{x}{z}q(\frac{x}{z},\mu^2)\left[\mathcal{C}_{T,L}^{c}+\mathcal{C}_{T,L}^{d}\right]\right\}, \nonumber
\end{eqnarray}
while the contribution from diagrams (e) and (f) is
\begin{equation} \label{eq:LOQCDef}
  \sigma^{\gamma^*p}_{T,L}=\sum_q \frac{4\pi^2\alpha e_q^2}{Q^2}\,\int_0^1\!\dif{X}\,\int_0^\infty\!\diff{k_t^2}\frac{\xB}{X}\,\frac{k_t^2}{Q^2}\;\sum_{\xi=\xi_\pm}\frac{\xB/\xi}{\left\lvert 1-b(X-\xB)/\xi \right\rvert}\,\frac{\alpha_S(k_t^2)}{2\pi}\,C_F\,Xq(X,\mu^2)\left[\mathcal{C}_{T,L}^{e}+\mathcal{C}_{T,L}^{f}\right],
\end{equation}
where the coefficients are
\begin{align}
  \mathcal{C}_T^{a} &= 1 - \frac{\beta\,\left( x + 2\,\xB\,z - 4\,x\,z - 2\,\xB\,z^2 + 4\,x\,z^2 \right) }{x\,\left( 1 - 2\,z + 2\,z^2 \right)},\qquad &
  \mathcal{C}_L^{a} &= \frac{8\,\beta\,(1-\beta)\,\xB\,\left( 1 - z \right) \,z}{x\,\left( 1 - 2\,z + 2\,z^2 \right) },\notag \\[5pt]
  \mathcal{C}_T^{b} &= A\left( x - 2\,\beta\,x - 2\,\beta\,\xB\,z - 2\,x\,z+ 4\,\beta\,x\,z \right), &
  \mathcal{C}_L^{b} &= 8\,A\,\beta\,\left( 1 - \beta \right)\,\xB\,z \,, \notag\\[5pt]
  \mathcal{C}_T^{c} &= 1 - \frac{\beta\,\left( x + \xB\,z - x\,z -\xB\,z^2 + 2\,x\,z^2 \right) }{x\,\left( 1 + z^2 \right) }, &
  \mathcal{C}_L^{c} &= \frac{4\,\beta\,(1-\beta)\,\xB\,\left( 1 - z \right)\,z}{x\,\left( 1 + z^2 \right) }, \notag\\[5pt]
  \mathcal{C}_T^{d} &= \frac{-\beta\,x\,z\,(1-z)}{\left( x-\xB\,z\right)\,\left( 1 + z^2 \right) }, &
  \mathcal{C}_L^{d} &= 0, \label{eq:LOQCDcoeffs}\\[5pt]
  \mathcal{C}_T^{e} &= \frac{-b\,\xi\,\left( \xi+\xB \right) }{\left( X - \xB \right) \,\left( X - \xB - \xi \right) \,\left( \xi + \left( 1 - b \right) \,\xB \right) }, &
  \mathcal{C}_L^{e} &= 0, \notag \\[5pt]
  \mathcal{C}_T^{f} &= \frac{\xi}{{\left( X - \xB \right) }^2}, &
  \mathcal{C}_L^{f} &= 0, \notag
\end{align}
where
\begin{equation}
  A = \beta\,(1-z) \Bigg/ \left[( x + \xB\,z - \beta\,\xB\,z - x\,z)( 1 - 2\,z + 2\,z^2)\right].
\end{equation}
Note that for high $E_T$ jet production in LO QCD there are no infrared singularities from either on-shell propagators or soft gluon emission.  We will take the factorisation scale to be $\mu=E_T=k_t$, in order to compare directly with the approach based on unintegrated partons.  The inclusive jet cross section calculated using \eqref{eq:LOQCDabcd} and \eqref{eq:LOQCDef} was found to be in excellent agreement with the LO QCD predictions of the \texttt{JetViP} \cite{Potter:1999gg} and \texttt{DISENT} \cite{Catani:1996vz} programs.

At this point it is an interesting check to take the DGLAP limit, so that we insert $\Theta(\mu-k_t)$ and take the limit $k_t\ll Q$, so that the only contributions come from the ladder-type diagrams of Fig.~\ref{fig:LOQCDcut} (a) and (c), and
\begin{equation}
  \sigma^{\gamma^*p}_T=\sum_q \frac{4\pi^2\alpha e_q^2}{Q^2}\,\int_x^1\!\dif{z}\,\int_0^{\mu^2}\!\diff{k_t^2}\;\frac{\alpha_S(k_t^2)}{2\pi}\;\left\{P_{qg}(z)\,\frac{x}{z}g(\frac{x}{z},\mu^2)+P_{qq}(z)\,\frac{x}{z}q(\frac{x}{z},\mu^2)\right\},
\end{equation}
with $x=\xB$ and $\sigma^{\gamma^*p}_L=0$.  At lowest order,
\begin{equation}
  F_2(\xB,\mu^2) = \frac{Q^2}{4\pi^2\alpha}\left(\sigma^{\gamma^*p}_T+\sigma^{\gamma^*p}_L\right)= \sum_q e_q^2 x\,q(x,\mu^2),
\end{equation}
leading to the well-known logarithmic scaling violation of $F_2$, or equivalently the ``real'' part of the DGLAP equation for the (integrated) quark distribution:
\begin{equation}
  \frac{\partial\,q(x,\mu^2)}{\partial \log(\mu^2)} = \frac{\alpha_S(\mu^2)}{2\pi}\,\int_x^1\!\diff{z}\;\left\{P_{qg}(z)\,g(\frac{x}{z},\mu^2)+P_{qq}(z)\,q(\frac{x}{z},\mu^2)\right\},
\end{equation}
where the conventional choice of scale is $\mu=Q$.  To obtain the ``virtual'' part of the DGLAP equation for the quark distribution, it is necessary to consider loop corrections to $\gamma^* q \to q$.  Of course, for high $E_T$ jet production, it is not appropriate to take the limit $k_t\ll Q$.

Let us anticipate how this calculation would be treated in terms of unintegrated partons, where we would want to factor out the emission with 4-momentum $j_2$ in Fig.~\ref{fig:LOQCDcut} (a) and (c) into the doubly-unintegrated quark distribution, $f_q(x,z,k_t^2,\mu^2)$.  For this to be possible, we must assume that $\mathcal{C}_T^{a}=1=\mathcal{C}_T^{c}$, and neglect all other contributions.  The diagrams in Fig.~\ref{fig:LOQCDcut} (d), (e) and (f) come from the subprocess $\gamma^* q \to qg$, where the gluon is radiated off the final quark line.  Such diagrams are strongly suppressed in an axial gluon gauge, due to one or more of the propagators having very large virtualities, and can be neglected.  Similarly, for the crossed quark box diagram of Fig.~\ref{fig:LOQCDcut} (b).  Numerically, the terms proportional to $\beta$ in diagrams (a) and (c) are found to be very small.  The one exception is the term proportional to $\beta$ in $\mathcal{C}_T^c$. This is negative and increasingly important as $z$ increases, that is, it is a destructive interference term.  In the case of our doubly-unintegrated quark, the same effect is obtained with an explicit constraint from angular ordering, so the term proportional to $\beta$ is redundant.

Ultimately, we will need to resort to explicit numerical comparison of $(z,k_t)$-factorisation with the conventional collinear factorisation approach in order to demonstrate the approximate equivalence of the two methods.

\subsection{$(z,k_t)$-factorisation approach at LO} \label{sec:xktLO}

With the new $(z,k_t)$-factorisation framework developed in Section \ref{sec:newfact} the LO diagram is simply $\gamma^*q\to q$, illustrated in Fig.~\ref{fig:LOjets}(a), where the incoming quark has 4-momentum $k=x\,p-\beta\,q^\prime+k_\perp$.  The partonic cross section contained in \eqref{eq:newktfact} is
\begin{equation}
  \dif\hat{\sigma}_{T,L}^{\gamma^*q}(x,z,k_t^2,\mu^2) = \dif\Phi^{\gamma^*q}\,\lvert\mathcal{M}_{T,L}^{\gamma^*q}\rvert^2\,/\,F^{\gamma^*q},
\end{equation}
where $F^{\gamma^*q} = 4x\,p\cdot q = 2x\,Q^2/\xB$.  Labelling the current jet by
\begin{equation}
  P=k+q=(x-\xB)\,p+(1-\beta)\,q^\prime+k_\perp,
\end{equation}
where $x\ge \xB$, the 1-body phase space is
\begin{eqnarray}
  \dif\Phi^{\gamma^*q} &=& (2\pi)^4\delta^{(4)}\,(k+q-P)\,\frac{\dif^4 P}{(2\pi)^3}\,\delta(P^2)\ =\ 2\pi\;\delta(\,P^2\,) \nonumber \\
  &=&  2\pi\;\frac{\xB}{Q^2}\sum_{i=\pm}\frac{1}{1-\xB\beta/x}\delta(\,x-x_i\,),
\end{eqnarray}
where $x_\pm$ is given by \eqref{eq:x_pm} with $r\equiv k_t^2/Q^2$.  Again, the condition $x\ge \xB$ means that only the $x=x_+$ solution contributes.  The rapidity of the current jet in the Breit frame is
\begin{equation} \label{eq:etaBJ}
  \eta^{{\rm Breit}}_P = \frac{1}{2}\log\frac{P^+}{P^-} = \frac{1}{2}\log\frac{x/\xB-1}{1-\beta}.
\end{equation}
\begin{figure}
  \begin{center}
    \includegraphics[width=0.35\textwidth]{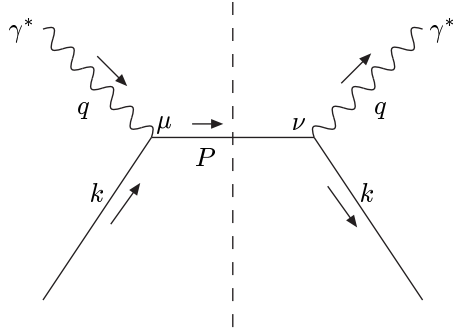}
    \caption{Cut diagram contributing at LO in the $(z,k_t)$-factorisation approach.\label{fig:LOzktcut}}
  \end{center}
\end{figure}
The squared matrix element, given by the cut diagram of Fig.~\ref{fig:LOzktcut}, is
\begin{equation} \label{eq:trace}
  \lvert\mathcal{M}_{T,L}^{\gamma^*q}\rvert^2 = \frac{1}{2}e^2e_q^2\;{\rm Tr}\left[\slashed{k}\,\gamma^\nu\,(\slashed{k}+\slashed{q})\,\gamma^\mu\,\right]\;\epsilon_\mu(q,\lambda)\,\epsilon_\nu^*(q,\lambda),
\end{equation}
where $\lambda$ is either $T$ or $L$.  We use the same formulae, \eqref{eq:sumepsT} and \eqref{eq:sumepsL}, to sum over the photon polarisations as before.

Note that our approach is not gauge invariant since we do not include the complete set of cut diagrams shown in Fig.~\ref{fig:LOQCDcut}.  Rather, we only keep the leading $\dif{k_t^2}/k_t^2$ term coming from Fig.~\ref{fig:LOQCDcut} (a) and (c).  We rely on using a physical gluon gauge where the neglected diagrams are suppressed.  We represent this approach by Fig.~\ref{fig:LOzktcut}, where the incoming quark is off-shell with virtuality $-k_t^2/(1-z)$.  Strictly speaking, the Ward identity \eqref{eq:WardID} does not apply to Fig.~\ref{fig:LOzktcut}.  For example, the $q_\mu$ term of the longitudinal photon polarisation vector \eqref{eq:epsL} gives rise to large cancellations between the contributions from Fig.~\ref{fig:LOQCDcut} (a) and (b) to ensure that the Ward identity is satisfied.  When the diagram of Fig.~\ref{fig:LOQCDcut}(b) is neglected, as in Fig.~\ref{fig:LOzktcut}, the $q_\mu$ term in $\epsilon_\mu(q,L)$ gives a much too large $\sigma_L$.  Therefore, we should not include the $q_\mu$ term in $\epsilon_\mu(q,L)$; this is equivalent to an appropriate choice for the photon gauge.

According to the prescription given in Section \ref{sec:newfact} we should only keep the leading $\dif{k_t^2}/k_t^2$ term in the squared matrix element and so terms explicitly of $\mathcal{O}(k_t^2/Q^2)$ should be neglected when calculating $\lvert\mathcal{M}_{T,L}^{\gamma^*q}\rvert^2$.  This amounts to the substitution $k=x\,p$ in the trace \eqref{eq:trace}, leading to
\begin{equation}
  \lvert\mathcal{M}_{T}^{\gamma^*q}\rvert^2 = 4\pi\alpha\,e_q^2\,Q^2\,\frac{x}{\xB}\qquad{\rm and}\qquad \lvert\mathcal{M}_{L}^{\gamma^*q}\rvert^2 = 0.
\end{equation}
The partonic cross sections are then
\begin{equation}
  \hat{\sigma}_T^{\gamma^*q}(x,z,k_t^2,\mu^2) = \frac{4\pi^2\alpha}{Q^2}\frac{\xB}{1-\xB\beta/x}\;\delta(x-x_+)e_q^2
  \qquad {\rm and}\qquad \hat{\sigma}_L^{\gamma^*q}(x,z,k_t^2,\mu^2) = 0.
\end{equation}
Inserting into \eqref{eq:newktfact} we obtain the hadronic cross section
\begin{equation} \label{eq:had}
  \sigma_T^{\gamma^*p} = \frac{4\pi^2\alpha}{Q^2}
  \int_x^1\!\dif{z}\,\int_0^{\infty}\!\diff{k_t^2}\;\frac{\xB/x}{1-\xB\beta/x}\;\sum_q e_q^2 f_q(x,z,k_t^2,\mu^2),
\end{equation}
with $x=x_+$.  Again, it is an interesting check to take the collinear limit, $k_t\ll Q$, so that we insert $\Theta(\mu-k_t)$ and take $\mu=Q$.  Then, $x\to \xB$, $\beta\to 0$ and by the normalisation condition \eqref{eq:norm} we recover the parton model prediction for the proton structure function $F_2=F_T+F_L$:
\begin{equation}
  F_2(\xB,Q^2) = \frac{Q^2}{4\pi^2\alpha}\;(\sigma_{T}^{\gamma^*p}+\sigma_{L}^{\gamma^*p}) = \sum_q e_q^2\,\xB\,q(\xB,Q^2).
\end{equation}
Alternatively, taking the limit $z\to 0$ of $x$ and $\beta$ in \eqref{eq:had}, then using the normalisation \eqref{eq:znorm0}, gives a $k_t$-factorisation prediction:
\begin{equation}
  F_2(\xB,\mu^2) = \int_0^{\infty}\!\diff{k_t^2}\;\frac{\xB}{x}\;\sum_q e_q^2 f_q(x,k_t^2,\mu^2),
\end{equation}
with $x = \xB (1+k_t^2/Q^2)$.

To test the assertion that the angular-ordering constraint mimics the major neglected terms in the LO QCD calculation of Section \ref{sec:LOQCD}, we can replace $P_{qg}(z)$ by $P_{qg}(z)\,(\mathcal{C}^a+\mathcal{C}^b)$ and $P_{qq}(z)$ by $P_{qq}(z)\,\mathcal{C}^c$ in the real part of the doubly-unintegrated quark, where the coefficients were given in \eqref{eq:LOQCDcoeffs}.  The inclusive jet cross section calculated in this manner, with separate coefficients for the $T$ and $L$ contributions, is found to be almost unchanged, providing evidence that the destructive interference terms in the conventional LO QCD calculation have much the same effect as an explicit angular-ordering constraint.

\subsection{Towards a NLO $(z,k_t)$-factorisation approach} \label{sec:NLOzkt}

\begin{figure}
  \begin{center}
    \includegraphics[width=0.8\textwidth]{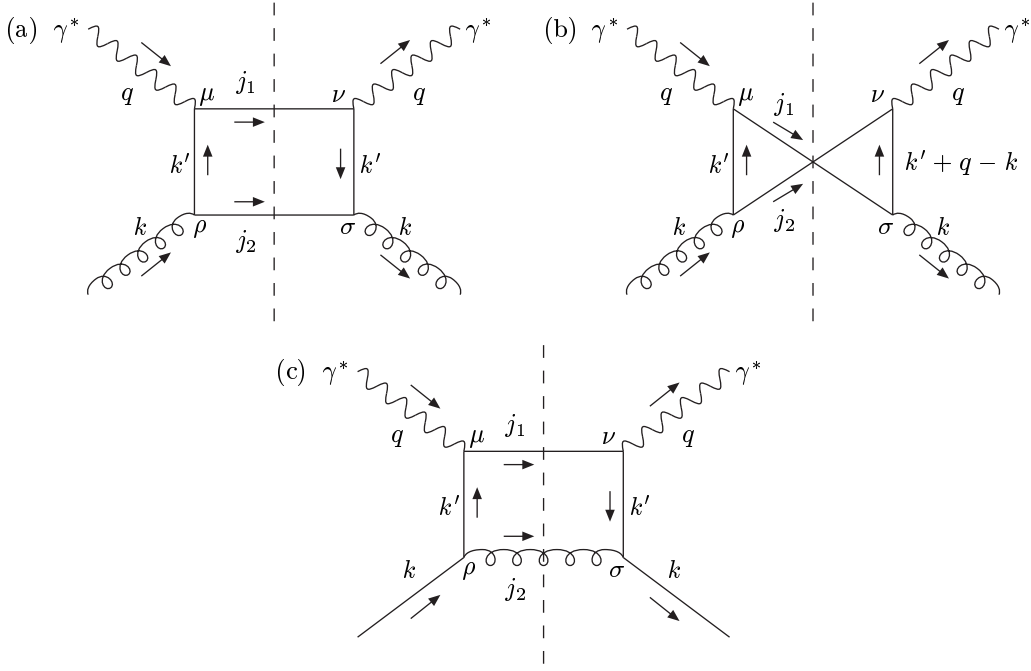}
    \caption{Cut diagrams contributing at ``NLO'' in the $(z,k_t)$-factorisation approach.\label{fig:NLOzktcut}}
  \end{center}
\end{figure}

It is beyond the scope of this work to perform a full NLO calculation within the framework of $(z,k_t)$-factorisation.  Rather, at this exploratory stage, we aim to produce a simplified description using the LO doubly-unintegrated partons and computing only the $\mathcal{O}(\alpha_S)$ diagrams expected to be dominant.  We do not want to include diagrams involving divergences which cannot be regulated by putting a cutoff on soft gluon emission from angular ordering.  The major loop corrections are already accounted for by the Sudakov form factor \eqref{eq:Sudakov}.  The diagram where a gluon is radiated from the final quark line is strongly suppressed in a physical gauge.  This leaves the cut diagrams of Fig.~\ref{fig:NLOzktcut} as the only contributions which should be included.  It is debatable whether or not the crossed box diagram of Fig.~\ref{fig:NLOzktcut}(b) should be included.  We choose to include it, although it gives only a relatively small contribution to the cross section.

All diagrams in Fig.~\ref{fig:NLOzktcut} have the same kinematics.  An initial parton, with 4-momentum $k = x\,p-\beta\,q^\prime+k_\perp$, splits to a quark with 4-momentum $k^\prime=x^\prime\,p-\beta^\prime\,q^\prime + k^\prime_\perp$, which goes on to interact with the photon.  The outgoing partons have 4-momentum $j_1 = k^\prime+q$ and $j_2 = k-k^\prime$ where $\xB \le x^\prime \le x \le 1$ and $0 \le \beta \le \beta^\prime  \le 1$.  Including the parton emitted in the last evolution step, the \texttt{KtJet} package \cite{Butterworth:2002xg} was used to cluster the three outgoing partons into jets using the inclusive $k_\perp$ algorithm in the Breit frame.  Note that the diagrams of Fig.~\ref{fig:NLOzktcut} naturally include the LO contribution of Fig.~\ref{fig:LOzktcut} in the limit that $k_t \ll k^\prime_t$.  Therefore, the LO contribution does not have to be added in explicitly.

We find that the 2-body phase space divided by the flux factor is given by
\begin{equation}
  \frac{\dif \Phi^{\gamma^*a}}{F^{\gamma^*a}} = \frac{\dif {k^\prime_t}^2}{16\pi x}\left(\frac{\xB}{Q^2}\right)^2
  \sum_{x^\prime=x^\prime_\pm} \left\lvert x-\xB\beta -(1-\beta)x^\prime-(x-\xB)\beta^\prime  \right\rvert^{-1},
\end{equation}
where $\beta^\prime  = \beta + (\xB R)/(x-x^\prime)$ and 
\begin{eqnarray}
  x^\prime_\pm\! & =\! & \frac{1}{2(1-\beta)}\biggl\{ x(1-\beta) + \xB(1-\beta-R) + \xB r^\prime \\
  && \ \left. \pm \sqrt{\left[\xB(1-\beta+R)-x(1-\beta)\right]^2+\xB r^\prime \left[\xB r^\prime-2 \left( x(1-\beta) -\xB(1-\beta-R)\right)\right]} \right\},\nonumber
\end{eqnarray}
with $r^\prime\equiv {k^\prime_t}^2/Q^2$ and $R\equiv {\lvert\vec{k_t}-\vec{k^\prime_t}\rvert}^2/Q^2$.

The cut diagrams representing the squared matrix elements are shown in Fig.~\ref{fig:NLOzktcut}.  Again, we write
\begin{equation}
  {\lvert \mathcal{M}_{T,L}^{\gamma^*a}\rvert}^2 = \frac{1}{2}\,e^2\,g^2\;M^{\mu\nu}\;\epsilon_\mu(q,\lambda)\epsilon_\nu^*(q,\lambda),
\end{equation}
where $\lambda$ is either $T$ or $L$ and the initial factor of $1/2$ is to average over the helicity of the incoming parton.  We take $e^2=4\pi\alpha(Q^2)$ and $g^2=4\pi\alpha_S(\mu_R^2)$, with $\mu_R={\rm max}(k_t,k^\prime_t)$. We have
\begin{eqnarray}
  {\rm (a)}\qquad M^{\mu\nu} &=& \left(\sum_q e_q^2\right) T_R\frac{1}{{k^\prime}^4}{\rm Tr}\left[\slashed{k}^\prime\gamma^\rho\slashed{j_2}\gamma^\sigma\slashed{k}^\prime\gamma^\nu\slashed{j_1}\gamma^\mu\right]d_{\rho\sigma}(k), \\
  {\rm (b)}\qquad M^{\mu\nu} &=& \left(\sum_q e_q^2\right) T_R\frac{1}{{k^\prime}^2}\frac{1}{({k^\prime}+q-k)^2}{\rm Tr}\left[\slashed{k}^\prime\gamma^\rho\slashed{j_2}\gamma^\nu(\slashed{k}^\prime+\slashed{q}-\slashed{k})\gamma^\sigma\slashed{j_1}\gamma^\mu\right]d_{\rho\sigma}(k), \\
  {\rm (c)}\qquad M^{\mu\nu} &=& e_q^2 C_F\frac{1}{{k^\prime}^4}{\rm Tr}\left[\slashed{k}^\prime\gamma^\rho\slashed{k}\gamma^\sigma\slashed{k}^\prime\gamma^\nu\slashed{j_1}\gamma^\mu\right]d_{\rho\sigma}(j_2).
\end{eqnarray}
In order to keep only the leading $\dif k_t^2/k_t^2$ term, we make the replacement $k\to x\,p$ in the numerator of these expressions, but not in the virtualities in the denominator.  Inserting the partonic cross sections into \eqref{eq:newktfact} we finally obtain
\begin{eqnarray}
  \sigma^{\gamma^*p}_{T,L}&=&\sum_q \frac{4\pi^2\alpha\,e_q^2}{Q^2}\,\left(\frac{\xB}{Q^2}\right)\,\int_0^1\!\dif{x}\,\int_x^1\!\dif{z}\,\int_0^\infty\!\diff{k_t^2}\,\int_0^\infty\!\dif{{k^\prime_t}^2}\,\frac{\xB}{x}\,\frac{\alpha_S(\mu_R^2)}{2\pi} \nonumber\\
  &&\quad\times\,\sum_{x^\prime=x^\prime_\pm} \left\lvert x-\xB\beta -(1-\beta)x^\prime-(x-\xB)\beta^\prime \right\rvert^{-1}\;\\
  &&\quad\times \left\{T_R\,f_g(x,z,k_t^2,\mu^2)\,\left[\mathcal{C}_{T,L}^{a}+\mathcal{C}_{T,L}^{b}\right]+C_F\,f_q(x,z,k_t^2,\mu^2)\,\mathcal{C}_{T,L}^{c}\right\},\nonumber
\end{eqnarray}
where the coefficients are
\begin{align}
  \mathcal{C}_T^{a} &= \frac{\left( 1 - 2\,\beta^\prime \,\left( 1 - \beta^\prime  \right) \right) \,x\, \left( x^\prime - \xB \right)  + \left( \beta^\prime \, \left( \xB - 2\,x^\prime \right)  + x^\prime \right) \, \left( \left( 1 - \beta^\prime  \right) \, \xB + \left( 2\,\beta^\prime  -1 \right)\,x^\prime \right) } {x\,{\left( \left( 1 - \beta^\prime  \right) \, \xB - x^\prime \right) }^2}, \notag \\[5pt]
  \mathcal{C}_L^{a} &= \frac{4\,\left( 1 - \beta^\prime \right) \,\beta^\prime \,\xB\, \left(x^\prime -\xB + \beta^\prime \,\left( x + \xB - 2\,x^\prime \right) \right) }{x\,{\left( \left( 1 - \beta^\prime  \right) \, \xB - x^\prime \right) }^2},\notag \\[5pt]
  \mathcal{C}_T^{b} &= \frac{\left( 1 - \beta^\prime  \right) \, \left( x^\prime - \xB \right) \, \left( \left( 1 - 2\,\beta^\prime  \right) \,x + 2\,\left( \beta^\prime \, \left( 2\,x^\prime -\xB\right) - x^\prime \right)  \right) } {x\,\left( x^\prime - \left( 1 - \beta^\prime  \right) \, \xB \right) \, \left( \left( 1 + \beta - \beta^\prime  \right) \,\left( x - x^\prime\right)+\left( 1 + \beta - \beta^\prime  + R \right)\,\xB\right) },\\[5pt]
  \mathcal{C}_L^{b} &= \frac{8\,{\left( 1 - \beta^\prime  \right) }^2\,\beta^\prime \,\xB\,\left( x^\prime - \xB \right) }{x\,\left( x^\prime - \left( 1 - \beta^\prime  \right) \,\xB \right) \, \left( \left( 1 + \beta - \beta^\prime  \right) \,\left( x -x^\prime\right) + \left( 1 + \beta - \beta^\prime  + R \right)\,\xB \right) },\notag \\[5pt]
  \mathcal{C}_T^{c} &= \frac{\left( 1 - 2\,\beta^\prime \, \left( 1 - \beta^\prime  \right)\right) \,x\, \left( x^\prime - \xB \right) + x^\prime\, \left( \left( 2\,\beta^\prime  - 1 \right) \, \xB + \left( 1 - 2\,{\beta^\prime }^2\right)\,x^\prime \right) } {\left( x - x^\prime \right) \, {\left( \left( 1 - \beta^\prime  \right) \,\xB - x^\prime \right) }^2},\notag \\[5pt]
  \mathcal{C}_L^{c} &= \frac{4\,\left( 1 - \beta^\prime  \right) \, {\beta^\prime }^2\,\xB}{{\left(\left( 1 - \beta^\prime  \right) \,\xB - x^\prime \right) }^2}.\notag
\end{align}
Inspection of the coefficient $\mathcal{C}_T^{c}$ reveals a pole at $z^\prime\equiv x^\prime/x=1$, corresponding to soft gluon emission.  We can regulate this singularity by appealing to angular ordering.  The rapidity of the gluon, with 4-momentum $j_2$, should be greater than the rapidity of the quark, with 4-momentum $j_1$:
\begin{equation}
  \eta^{{\rm Breit}}_{j_2} > \eta^{{\rm Breit}}_{j_1} \quad\iff\quad z^\prime < \frac{\mu^\prime}{\mu^\prime+\left\lvert\vec{k_t}-\vec{k^\prime_t} \right\rvert}, \qquad{\rm with}\qquad \mu^\prime \equiv Q\frac{x^\prime}{\xB}\sqrt{\frac{1-\beta^\prime }{x^\prime/\xB-1}}.
\end{equation}
This condition applies only to the diagram where a quark radiates a gluon, Fig.~\ref{fig:NLOzktcut}(c), but not to the diagrams where a gluon radiates a quark, Fig.~\ref{fig:NLOzktcut} (a) and (b).

\subsection{An estimate of the NNLO contribution}

\begin{figure}
  \begin{center}
    \includegraphics[width=0.8\textwidth]{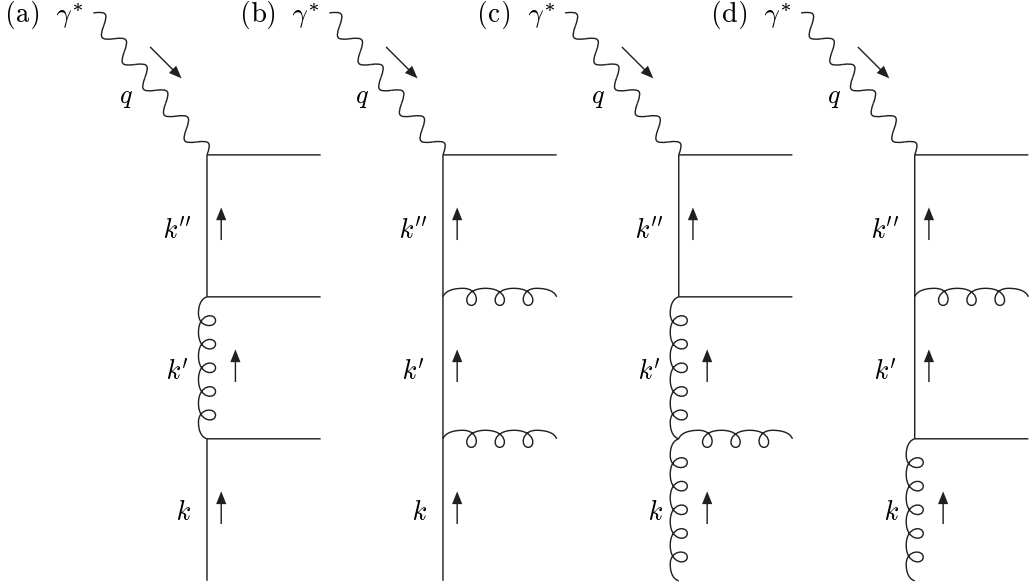}
    \caption{Feynman diagrams contributing at ``NNLO'' in the $(z,k_t)$-factorisation approach.\label{fig:NNLOzkt}}
  \end{center}
\end{figure}

The next-to-next-to-leading order diagrams have not yet been calculated in the collinear approximation (NNLO QCD).  As we explain later, the ``NLO'' calculation of Section \ref{sec:NLOzkt} gives reasonable agreement with conventional NLO QCD.  It is possible that a simplified ``NNLO'' $(z,k_t)$-factorisation calculation may provide an estimate of whether the NNLO QCD corrections are likely to be important, especially at low $E_T$ and low $Q^2$ in the forward region, where there is a discrepancy between NLO QCD and the data.

The four contributing diagrams, all of which have the same kinematics (phase space), are shown in Fig.~\ref{fig:NNLOzkt}.  Diagrams (a) and (b) are the doubly-unintegrated quark contribution, while diagrams (c) and (d) are the doubly-unintegrated gluon contribution.  Encouraged by the fact that the crossed quark box of Fig.~\ref{fig:NLOzktcut}(b) gave only a small contribution, we may neglect the interference cut graphs arising from Fig.~\ref{fig:NNLOzkt} as a first approximation, leaving only four squared matrix elements to be calculated.

Our simplified approach provides an approximation of QCD, in which only ladder-type diagrams remain.  The soft gluon singularities are regulated by angular ordering.  There are no infrared singularities remaining.  We can add an arbitrary number of rungs to the ladder and the answer will be finite.  However, with more rungs, the number of neglected interference terms grows; it is likely that the approximate treatment of these terms by imposing angular-ordering constraints will spoil the accuracy of the method if too many rungs are added.

\section{Description of HERA inclusive jet production data} \label{sec:cfdata}

HERA data are available for inclusive jet production in DIS. We may therefore check how well the simpler $(z,k_t)$-factorisation approach is able to reproduce the conventional collinear factorisation approach, and at the same time see how well these calculations describe the data.

Recall from Section~\ref{sec:jetprod} that at LO the $(z,k_t)$-factorisation approach is based on the simple $\gamma^*q\to q$ subprocess driven by the doubly-unintegrated quark distribution, $f_q(x,z,k_t^2,\mu^2)$, retaining the full kinematics. On the other hand, in the LO QCD description the subprocesses are $\gamma^*g\to q\bar q$ and $\gamma^* q\to gq$ evaluated with collinear kinematics and conventional integrated distributions, $g(x,Q^2)$ and $q(x,Q^2)$.

\subsection{Comparison with ZEUS data at high $Q^2$}

We now compare our predictions to the experimental data obtained by the ZEUS Collaboration \cite{Chekanov:2002be}.  This data was taken during 1996 and 1997, when HERA collided protons of energy $E_p=820$ GeV with positrons of energy $E_e=27.5$ GeV at a centre-of-mass energy of $\sqrt{s} = \sqrt{4E_p E_e}\simeq 300$ GeV.  Rather than make cuts on the variable $y=Q^2/(\xB s)$, ZEUS make cuts on $\cos \gamma$, one of the angles used in reconstructing the kinematical variables using the double-angle method, where
\begin{equation}
  \cos\gamma = \frac{\xB(1-y)E_p-yE_e}{\xB(1-y)E_p+yE_e}.
\end{equation}
In the parton model, the angle $\gamma$ corresponds to the direction of the scattered quark.  In \eqref{eq:jetcross} we therefore set $y_{\rm min}=0$ and $y_{\rm max}=1$ and demand instead that $\cos \gamma$ satisfies the ZEUS experimental cuts, $-0.7<\cos\gamma<0.5$.

\begin{figure}
  \begin{center}
    \includegraphics[clip,width=0.9\textwidth]{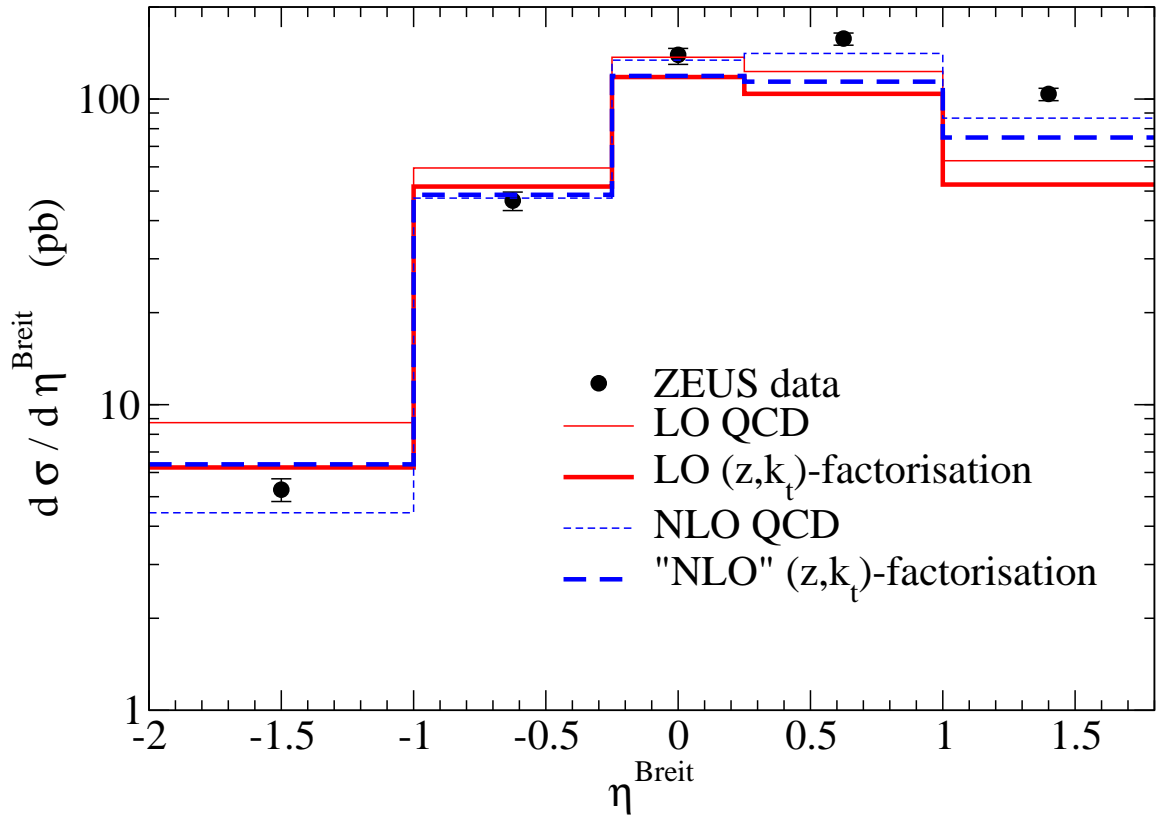}
    \caption{Comparison with ZEUS inclusive jet production data~\cite{Chekanov:2002be} at high $Q^2$. The feint and bold lines correspond, respectively, to the predictions of the conventional QCD approach and the $(z,k_t)$-factorisation approach based on doubly-unintegrated parton distributions. \label{fig:ZEUSplot}}
  \end{center}
\end{figure}

In Fig.~\ref{fig:ZEUSplot} we show the rapidity distribution, $\dif\sigma/\dif\eta^{\rm Breit}$, integrated over $Q^2$ from 125 to $10^5$ GeV$^2$ and over $E_T$ from 8 to 100 GeV.  The parton-to-hadron correction factors given in Table 3 of the ZEUS paper \cite{Chekanov:2002be} have been applied to the theory predictions.  For the results presented, we used the MRST2001 LO parton set \cite{Martin:2002dr} as input.  The NLO QCD predictions have been taken from the plot in Fig.~3b) of \cite{Chekanov:2002be}; these were obtained with the \texttt{DISENT} program \cite{Catani:1996vz} using MRST99 partons \cite{Martin:1999ww}, a renormalisation scale of $E_T$ and a factorisation scale of $Q$.  The statistical, systematic and jet-energy-scale uncertainties have been added in quadrature to estimate the total experimental uncertainty.  All the theory predictions give a reasonably good description of the data.  The NLO predictions generally give a slightly better description than the LO predictions.  For the $(z,k_t)$-factorisation approach, the ``NLO'' corrections are only significant in the forward region.

In order to verify that the extra $z$ convolution of $(z,k_t)$-factorisation with respect to $k_t$-factorisation is important, we also repeated the calculation taking the limit $z\to 0$ in the partonic cross section.  (Usually, $k_t$-factorisation is only applied using the unintegrated gluon, whereas here we also include the unintegrated quark.)  The parton emitted in the last evolution step then goes in the proton direction and is not counted in the inclusive jet cross section.  In general, the predictions are much worse, even in the current jet region, providing evidence that the extra $z$ convolution of our method is important.

\subsection{Comparison with H1 data at low $Q^2$}

The H1 Collaboration have measured the inclusive jet cross section in DIS at high $Q^2$ \cite{Adloff:2000tq} and at low $Q^2$ \cite{Adloff:2002ew}.  Here, we focus on the latter, where $Q^2 = 5$ to $100$ GeV$^2$.  In this region, the NLO QCD corrections to LO QCD are larger than at high $Q^2$, and the advantages of our approach become more apparent.  Again, this data was taken during 1996 and 1997.

The H1 Collaboration use the electron method to reconstruct the kinematical variables, so cuts are imposed directly on the variable $y$, namely $0.2< y < 0.6$.  We therefore set $y_{\rm min}=0.2$ and $y_{\rm max}=0.6$ in \eqref{eq:jetcross}. Also, H1 present their data in rapidity bins in the \textsc{lab} frame rather than the Breit frame. It can be shown that the rapidity in the $\textsc{lab}$ frame is
\begin{equation} \label{eq:etalab}
\eta_j^\textsc{lab} = \frac{1}{2}\log\left[\left(  4E_pE_e\frac{  \xB}{Q^2}\left(\frac{a_j}{b_j}+  \xB\right)-  \xB -
\frac{2  \xB}{b_j Q^2}e_tj_t\cos\phi_{ej}\right)\frac{E_p}{E_e}\right]\,,
\end{equation}
where, in the Breit frame, the initial positron has transverse momentum squared of
\begin{equation}
e_t^2 = 4E_pE_e \xB(4E_pE_e\frac{\xB}{Q^2}-1)
\end{equation}
and the 4-momentum of the outgoing jet has been written in the form
\begin{equation}
  j = a_j\,p + b_j\,q^\prime + j_\perp.
\end{equation}
It is necessary to average the cross section over the azimuthal angle $\phi_{ej}$ between the positron and the outgoing jet in the transverse plane.  For the ``NLO'' $(z,k_t)$-factorisation calculation, the jet 4-momenta are not necessarily the same as the 4-momenta of the outgoing partons.   It is necessary to pass the 4-momenta through a jet algorithm.  Rather than use \eqref{eq:etalab} to determine $\eta^\textsc{lab}$, which would require an additional azimuthal averaging, it is simpler to explicitly transform the 4-momenta from the Breit to the \textsc{lab} frame, then calculate the rapidity of the resultant 4-momenta.

\begin{figure}
  \begin{center}
    \includegraphics[clip,width=1.0\textwidth]{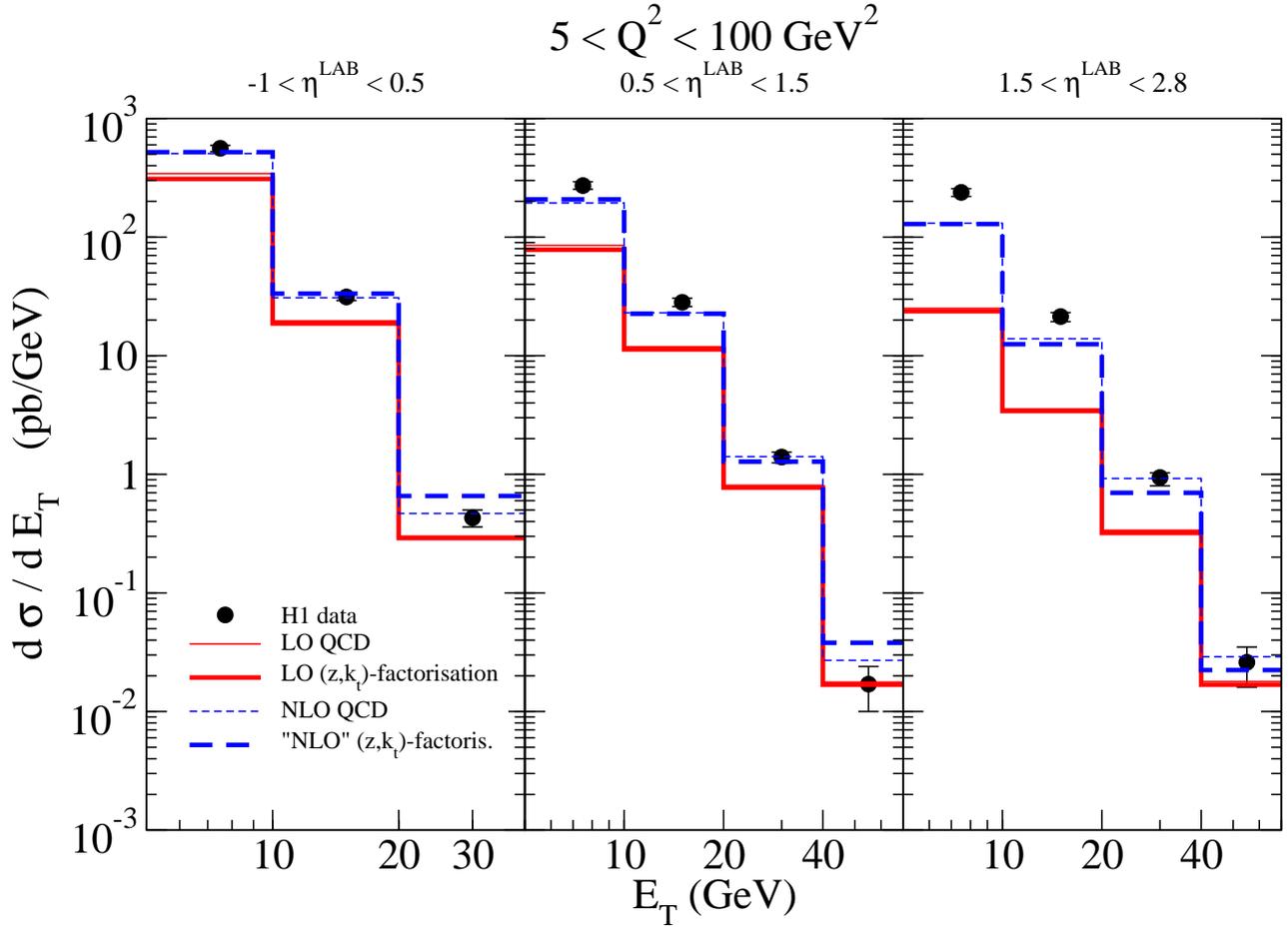}
    \caption{Comparison with H1 inclusive jet production data \cite{Adloff:2002ew} at low $Q^2$. The predictions of the $(z,k_t)$-factorisation approach based on doubly-unintegrated partons (which is much simpler to implement) are in good agreement with the conventional QCD approach.  In some bins the predictions of the latter approach are hidden beneath the bold lines of the $(z,k_t)$-factorisation approach, at the respective order. \label{fig:H1plot}}
  \end{center}
\end{figure}

In Fig.~\ref{fig:H1plot} we show $\dif\sigma/\dif E_T$ integrated over $Q^2$ between 5 and 100 GeV$^2$ in three rapidity intervals.  For the results presented, we used the MRST2001 LO parton set \cite{Martin:2002dr} as input.  The NLO QCD predictions have been taken from the plot in Fig.~1 of the H1 paper \cite{Adloff:2002ew}; these were obtained with the \texttt{DISENT} program \cite{Catani:1996vz} using CTEQ5M partons \cite{Lai:1999wy}, a renormalisation scale of $E_T$ and a factorisation scale of $Q$.  The hadronisation correction factors used in \cite{Adloff:2002ew} have been applied to all the theory predictions.  The statistical and systematic uncertainties have been added in quadrature to estimate the total experimental uncertainty.

The LO $(z,k_t)$-factorisation calculation is in excellent agreement with conventional LO QCD, but neither describe the data well, especially in the forward rapidity region.  The ``NLO'' $(z,k_t)$-factorisation calculation is in very good agreement with conventional NLO QCD, although the agreement gets slightly worse as $E_T$ increases.\footnote{In two bins the ``NLO'' $(z,k_t)$-factorisation predictions are significantly higher than the NLO QCD predictions.  This is due to the jet algorithm applied, which increases the ``NLO'' $(z,k_t)$-factorisation predictions by more than a factor of two in these two bins only, compared to the result when no jet algorithm is applied.}  Deviations of the data from NLO QCD are seen only at small $E_T$ in the forward region.  Here, the NLO corrections are quite large and it is likely that NNLO corrections or resolved virtual photon contributions are important in this region.  Again, taking the limit $z\to 0$ makes the $(z,k_t)$-factorisation predictions much worse, showing that it is important to keep the precise kinematics.

\section{Conclusions} \label{sec:conclusions}

We have presented a method for determining unintegrated parton distributions, $f_a(x,k_t^2,\mu^2)$, from the conventional (integrated) parton distributions, by considering the last DGLAP evolution step separately, and imposing angular-ordering constraints on gluon emission.  To include the precise kinematics in the hard subprocess initiated by the final parton in the evolution ladder, it is necessary to consider \emph{doubly}-unintegrated parton distributions, $f_a(x,z,k_t^2,\mu^2)$.  We gave a prescription, called $(z,k_t)$-factorisation, for the computation of cross sections using these distributions.  This prescription is a natural generalisation of the $k_t$-factorisation approach.

We used $(z,k_t)$-factorisation to estimate the cross section for inclusive jet production at HERA at lowest order.  Using the same LO doubly-unintegrated distributions, we then carried out a ``NLO'' calculation which included the dominant Feynman diagrams with the soft gluon singularities being regulated by angular ordering.

We showed that at $\mathcal{O}(\alpha_S^0)$ the predictions of the approach based on doubly-unintegrated partons, with exact kinematics, are close to the conventional LO QCD calculation at $\mathcal{O}(\alpha_S^1)$. The relative simplicity of the former approach is shown schematically in Fig.~\ref{fig:LOjets}.  Similarly, at $\mathcal{O}(\alpha_S^1)$ the predictions of the approach based on doubly-unintegrated partons are close to the conventional NLO QCD calculation at $\mathcal{O}(\alpha_S^2)$.

It was seen that the NLO corrections are large in the forward region at low $E_T$ and low $Q^2$ where the agreement with the data is poor.  It is possible that the simplified $(z,k_t)$-factorisation approach might help to evaluate the r\^{o}le of the NNLO contribution.  Alternatively, the resolved photon contribution is known to be important in the regime where $E_T$ is much greater than $Q$.  It would be better to calculate the resolved photon contribution in terms of the doubly-unintegrated parton distributions of the photon.\footnote{In Ref.~\cite{Motyka:2002ww}, for example, the KMR prescription was applied to obtain the unintegrated gluon distribution of the photon.}

The logical next step would be to show that doubly-unintegrated distributions can be applied to $pp$ and $p\bar{p}$ collisions.  The simplest calculation would be the transverse momentum distribution of produced $W$ and $Z$ bosons.

We conclude that by reorganising the perturbative expansion in $\alpha_S$ to keep only the most important terms, our method provides a simple but effective way of estimating exclusive (and inclusive) observables.

\section*{Acknowledgements}

We thank G\"unter Grindhammer, Thomas Sch\"orner-Sadenius and Markus Wobisch for providing supplementary H1 data and NLO QCD predictions from the \texttt{DISENT} program.  We also thank the UK Particle Physics and Astronomy Research Council and the Russian Fund for Fundamental Research (grants 01-02-17095 and SS-1124.2003.2) for financial support.

\end{document}